\documentclass{jpsj3}
\usepackage{comment}
\usepackage{amsmath}
\usepackage{bm}
\newcommand{\eref}[1]{(\ref{#1})}
\newcommand{\fref}[1]{Fig. \ref{#1}}
\newcommand{\kdotp}{{\boldsymbol k}\cdot{\boldsymbol p}}
\newcommand{\kp}{K^+}
\newcommand{\bkp}{{\boldsymbol K}^+}
\newcommand{\km}{K^-}
\newcommand{\bkm}{{\boldsymbol K}^-}
\newcommand{\kpm}{K^\pm}

\newcommand{\rr}{{\boldsymbol r}}

\newcommand{\Fap}{F_{\rm A}^+}

\newcommand{\Fam}{F_{\rm A}^-}
\newcommand{\Fbm}{F_{\rm B}^-}
\newcommand{\Fapm}{F_{\rm A}^\pm}
\newcommand{\Fbpm}{F_{\rm B}^\pm}
\newcommand{\psia}{\psi_{\rm A}({\boldsymbol r})}
\newcommand{\psib}{\psi_{\rm B}({\boldsymbol r})}

\def\runauthor{Yuji {\sc Shimomura} \textit{et al}.}
\title{%
Electronic States and Local Density of States in Graphene\\
with a Corner Edge Structure
}

\author{%
Yuji {\sc Shimomura}, Yositake {\sc Takane}, and
Katsunori {\sc Wakabayashi}$^{1,2}$
}

\inst{%
Department of Quantum Matter, Graduate School of Advanced Sciences of Matter,\\
Hiroshima University, Higashihiroshima, Hiroshima 739-8530, Japan\\
${}^{1}$International Center for Materials Nanoarchitectonics (MANA),
National Institute for Materials Science (NIMS), Namiki, Tsukuba 305-0044, Japan\\
${}^{2}$PRESTO, Japan Science and Technology Agency, Kawaguchi 332-0012, Japan
}

\recdate{ \hspace{50mm} }

\abst{%
We study electronic states of semi-infinite graphene
with a corner edge, focusing on the stability of edge localized states
at zero energy.
The $60^{\circ}$, $90^{\circ}$, $120^{\circ}$ and $150^{\circ}$ corner edges are examined.
The $60^{\circ}$ and $120^{\circ}$ corner edges consist of two zigzag edges,
while $90^{\circ}$ and $150^{\circ}$ corner edges consist of one zigzag edge and one armchair edge.
We numerically obtain the local density of states (LDOS)
on the basis of a nearest-neighbor tight-binding model
by using Haydock's recursion method.
We show that edge localized states appear along a zigzag edge
of each corner edge structure
except for the $120^{\circ}$ case.
To provide insight into this behavior,
we analyze electronic states at zero energy
within the framework of an effective mass equation.
The result of this analysis is consistent with the behavior of the LDOS.
}

\kword{%
graphene corner edge, localized state, zigzag edge, armchair edge, tunneling spectroscopy
}
\begin{document}
\maketitle
\section{Introduction}

The realization of a monolayer graphene
sheet~\cite{novoselov,novoselov.2005} has triggered
extensive studies on its unusual electronic properties arising from
the two-dimensional honeycomb structure of carbon atoms.~\cite{castro_neto}
Since the unit cell of the honeycomb lattice contains two nonequivalent
sites which form two sublattices A and B,
the low-energy electronic states of graphene near the Fermi energy 
are described by a $2 \times 2$ matrix form which is equivalent to the massless Dirac equation.~\cite{mcclure}
Thus, electrons in graphene are called massless Dirac fermions.
The band structure of massless Dirac fermions has a unique character,
since they have linear energy dispersion in the vicinity of two nonequivalent
symmetric points, called $K^{+}$ and $K^{-}$ points, in the Brillouin zone,
where the conduction and valence bands conically touch.~\cite{wallace}
This structure is called Dirac cone.
We hereafter set the electron energy at the band touching point
as $\varepsilon = 0$.
The unique energy band structure provide 
a number of intriguing physical properties such as 
the half-integer quantum Hall effect,~\cite{novoselov.2005,zhang}
the absence of backward scattering associated with the Berry's phase by
$\pi$~\cite{ando} and Klein tunneling.~\cite{klein}

The presence of edges makes an strong impact on the Dirac fermions in graphene near the Fermi energy.
As stressed by Fujita \textit{et al}., the electronic states near the
graphene edge strongly depends on its edge orientation.~\cite{fujita}
Typical straight edges of graphene are classified into two structures:
one is zigzag (zz) edge and the other is armchair (ac) edge.
Fujita \textit{et al}. analyzed electronic states in graphene with an infinitely
long straight edge on the basis of a nearest-neighbor tight-binding model,
and showed that highly degenerate edge localized states
appear at $\varepsilon = 0$ along a zz edge.~\cite{fujita}
These states at $\varepsilon = 0$ result in a sharp zero-energy peak
structure in the local density of states (LDOS)
near a straight zz edge of graphene.
The edge localized states have a characteristic feature that
their probability amplitude is finite only on one sublattice
including edge sites and completely vanishes on the other sublattice.
No such localized states appear along an ac edge.
The presence of edge localized states along a zz edge
has been confirmed by using scanning tunneling microscopy and scanning tunneling
spectroscopy.~\cite{kobayashi,niimi}

\begin{figure}
	\begin{center}
		\includegraphics[scale=0.6]{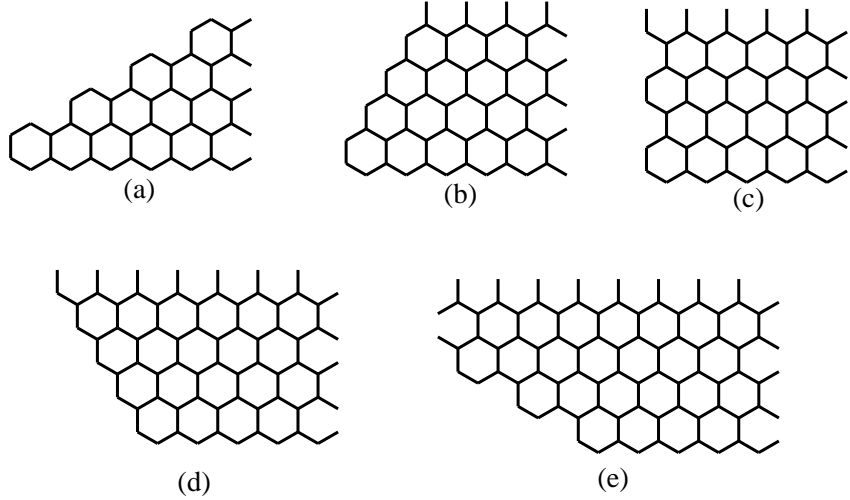}
	\end{center}
         \caption{Typical corner edge structures with coner angles
 of (a)30$^\circ$, (b)60$^\circ$, (c)90$^\circ$, (d)120$^\circ$ and
 (e)150$^\circ$. }
	\label{typicalcorner}
\end{figure}
Theoretically, the presence or absence of zero-energy localized states
has been well understood for infinitely long straight edges.
However, actual edges of graphene samples are never straight nor
infinitely long, and are much more complex than ideal ones.
An actual edge line consists of several zz and/or ac segments,
and a corner edge inevitably appears at the boundary of two adjacent segments.
Typical corner edge structures are shown in \fref{typicalcorner}.
Hereafter each corner edge is referred to according to its corner angle.
The $30^{\circ}$, $90^{\circ}$, and $150^{\circ}$ corner edges consist
of one zz edge and one ac edge, while the $60^{\circ}$ and $120^{\circ}$
corner edges consist of two zz edges.
There arises a natural question: Do edge localized states exist
at $\varepsilon = 0$ along a bent edge of these corner edge structures?
In this paper we study electronic states in the corner edge structures
to answer this question.
We adopt a nearest-neighbor tight-binding model and numerically obtain
the LDOS by using Haydock's recursion method.
We find that edge localized states appear along a zz edge of each corner edge
structure with an exception of the $120^{\circ}$ corner edge.
In the $120^{\circ}$ case, edge localized states locally disappear
near the corner but emerge with increasing the distance from the corner.
To provide insight into these unexpected behaviors,
we analyze electronic states at $\varepsilon = 0$
within the framework of an effective mass equation.
The result of this analysis is consistent with the behavior
of the LDOS.

\section{Formulations for Numerical Analysis}

\subsection{Model of graphene corner edges}

We describe
$\pi$ electrons in graphene with a corner edge structure by using a
tight-binding model on a honeycomb lattice.
The Hamiltonian of this model is represented as　
\begin{equation}
	H=
	-t\sum_{<i,j>}|i \rangle\langle j|
	+\sum_{i}w_i|i \rangle \langle i|,
	\label{hamiltonian}
\end{equation}
where $t$ is the nearest neighbor hopping integral 
and $w_i$ is a site-dependent potential.
If $w_i=0$ for any $i$, this model corresponds to a bulk graphene sheet.
The site-dependent potential $w_i$ is
introduced for a techinical reason.
For practical application of our numerical approach,
it is convenient to treat a lattice system being infinite in 
both the longitudinal and transverse directions.
However, such a system contains lattice sites which are irrelevant for 
a corner edge structure.
To model a corner edge on this infinite system,
we put a large on-site potential on each irrelevant site to prevent electrons arriving on it.
Therefore, we set $w_i=w$ with a sufficiently large $w$ 
if the $i$th site is irrelevent for a corner edge structure while $w_i=0$ otherwise.

We consider four corner edges
having corner angles differ from each other.
The angles are
60$^\circ$, 90$^\circ$, 120$^\circ$ and 150$^\circ$.
We particularly focus on corner edges including one or two zz edges.

\subsection{Haydock's recursion method}
The LDOS can be calculated with Haydock's recursion method~\cite{Hay,kelly,rec1,rec2}
which in applicable to systems having no translational symmetry
such as graphene with a corner edge.
By applying this method, we can obtain the LDOS at an arbitrary site.

We outline the method to obtain the LDOS at an $i$th site. 
To start with, we transform our model to a one-dimensional chain model.
We first introduce the coefficient $a_0$ given by
\begin{align}
	a_0&=\langle l_0|H|l_0\rangle
	\label{a_0}
\end{align}
with $|l_0\rangle\equiv|i\rangle$,
and define $|l_1\rangle$ and $b_1$ in terms of
\begin{align}
	b_{1}|l_1\rangle&=(H-a_0)|l_0\rangle
	\label{rec_eqat0}
\end{align}
with $\langle l_1|l_1\rangle\equiv1$.
The coefficient $b_1$ is obtained as
\begin{align}
	b_1=\sqrt{\langle l_0|(H-a_0)(H-a_0)|l_0\rangle}.
	\label{b_1}
\end{align}
We next introduce $a_1$ given by
\begin{align}
	a_1&=\langle l_1|H|l_1\rangle,
	\label{a_1}
\end{align}
and define $|l_2\rangle$ and $b_2$ in terms of
\begin{align}
	b_{2}|l_2\rangle&=(H-a_1)|l_1\rangle-b_1|l_0\rangle
	\label{rec_eqat1}
\end{align}
with $\langle l_2|l_2\rangle\equiv1$.
The coefficient $b_2$ is obtained as
\begin{align}
	b_2=
	\sqrt{ \left\{ \langle l_1|(H-a_1)-\langle l_0|b_1\right\}
	\left\{(H-a_1)|l_1\rangle-b_1|l_0\rangle \right\}}.
	\label{b_2}
\end{align}
Repeating this $n$ times, we obtain
\begin{align}
	b_{n+1}|l_{n+1}\rangle&=(H-a_n)|l_n\rangle-b_n|l_{n-1}\rangle,
	\label{rec_eqatn}
\end{align}
with
\begin{align}
	a_n&=\langle l_n|H|l_n\rangle,
	\label{a_n}
\end{align}
\begin{align}
	b_{n+1}=
	\sqrt{ \left\{ \langle l_n|(H-a_n)-\langle l_{n-1}|b_n\right\}
	\left\{(H-a_n)|l_n\rangle-b_n|l_{n-1}\rangle \right\}}.
	\label{b_n+1}
\end{align}

This manipulation with the reccurence equation, eq. \eref{rec_eqatn},
is equivalent to a transformation of the original electron system 
to a one-dimentinal chain model.
$\{|l_0\rangle, |l_1\rangle, |l_2\rangle, \ldots\}$ stands for
the orthonormal basis set of the chain model.
Here, $|l_n\rangle$ involves 
neighboring sites of $|i\rangle$ up to the $n$th nearest neighbors.
On this basis, 
$H$ can be rewritten
with real coefficients $\{a_0,a_1,\ldots\}$ and $\{b_1,b_2,\ldots\}$
as a tridiagonal matrix
\begin{equation}
	H=
	\begin{pmatrix}
		a_0&b_1&&\\
		b_1&a_1&b_2&\\
		&b_2&a_2&b_3\\
		&&b_3&a_3\\
		&&&&\ddots
	\end{pmatrix}.
	\label{Hm}
\end{equation}
	
With the coefficients $\{a_0,a_1,\ldots\}$ and $\{b_1,b_2,\ldots\}$,
the Green's function $G_i(E)$ for the $i$th site can be represented as a continued fraction, 
\begin{eqnarray}
	G_i(E)=\frac{1}{E-a_0-\frac{b_1^2}{E-a_1-\frac{b_2^2}{\ldots}}}.
	\label{G}
\end{eqnarray}
Practically,
we need to terminate this continued fraction
at a sufficiently large $n$.
If it is terminated at $n=N$,
we obtain the approximate expression of $G_i(E)$
as
\begin{eqnarray}
	G_i(E)=\frac{1}{E-a_0-\frac{b_1^2}{E-a_1-\frac{b_2^2}{\frac{\ldots}{E-a_N-t(E)}}}},
	\label{Gt}
\end{eqnarray}
where
\begin{equation}
	t(E)=\frac{E-a_N}{2b_N^2}\left[1-\{1-
	\frac{4b_N^2}{(E-a_N)^2}\}^{\frac{1}{2}}\right].
	\label{t}
\end{equation}
$G_i(E)$ gives the LDOS at the $i$th site in terms of the relation
\begin{equation}
	N_i(E)=\frac{1}{\pi}{\rm Im}G_i(E-{\rm i}\delta),
	\label{N(E)}
\end{equation}
where $\delta$ is a positive infinitesimal.
In actual numerical calculations,
we treat $\delta$ as a sufficiently small but finite constant.

\section{The LDOS}
\subsection{The LDOS in the presence of a single edge}
\begin{figure}[t]
	\begin{center}
		(a)\\
		\includegraphics[scale=0.6]{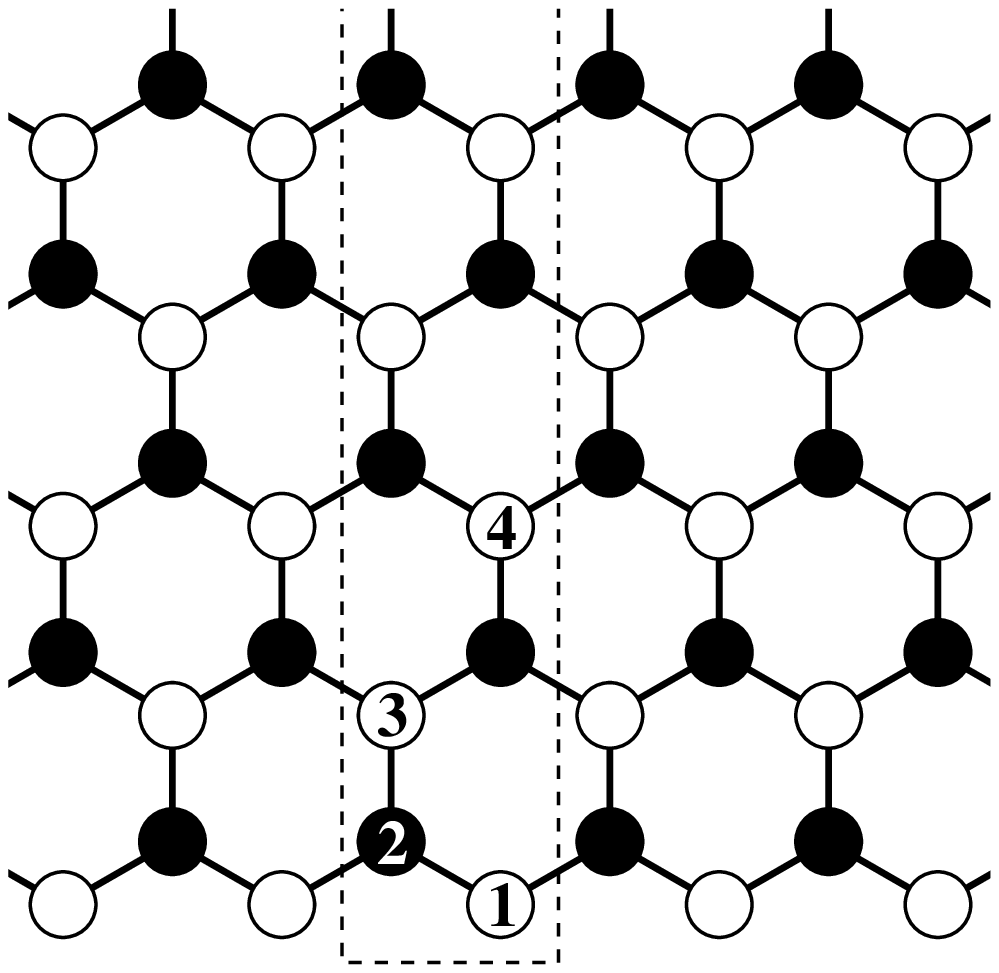}
	\end{center}
	\begin{minipage}[t]{0.495\textwidth}
		\begin{center}
			(b)\\
			1\\
			\includegraphics[scale=0.17]
			{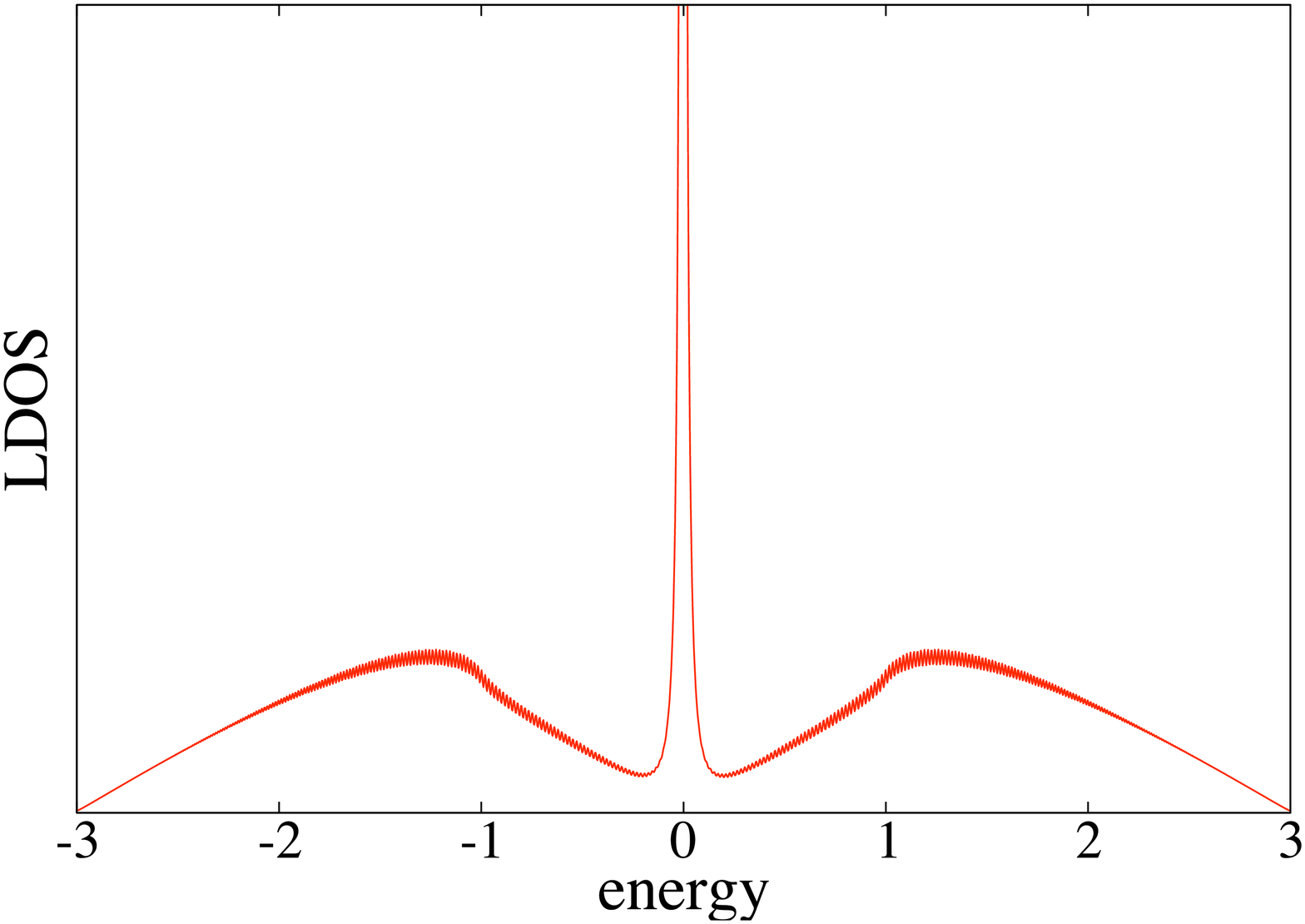}\\
		\end{center}
	\end{minipage}
	\begin{minipage}[t]{0.495\textwidth}
		\begin{center}
			(c)\\
			2\\
			\includegraphics[scale=0.17]
			{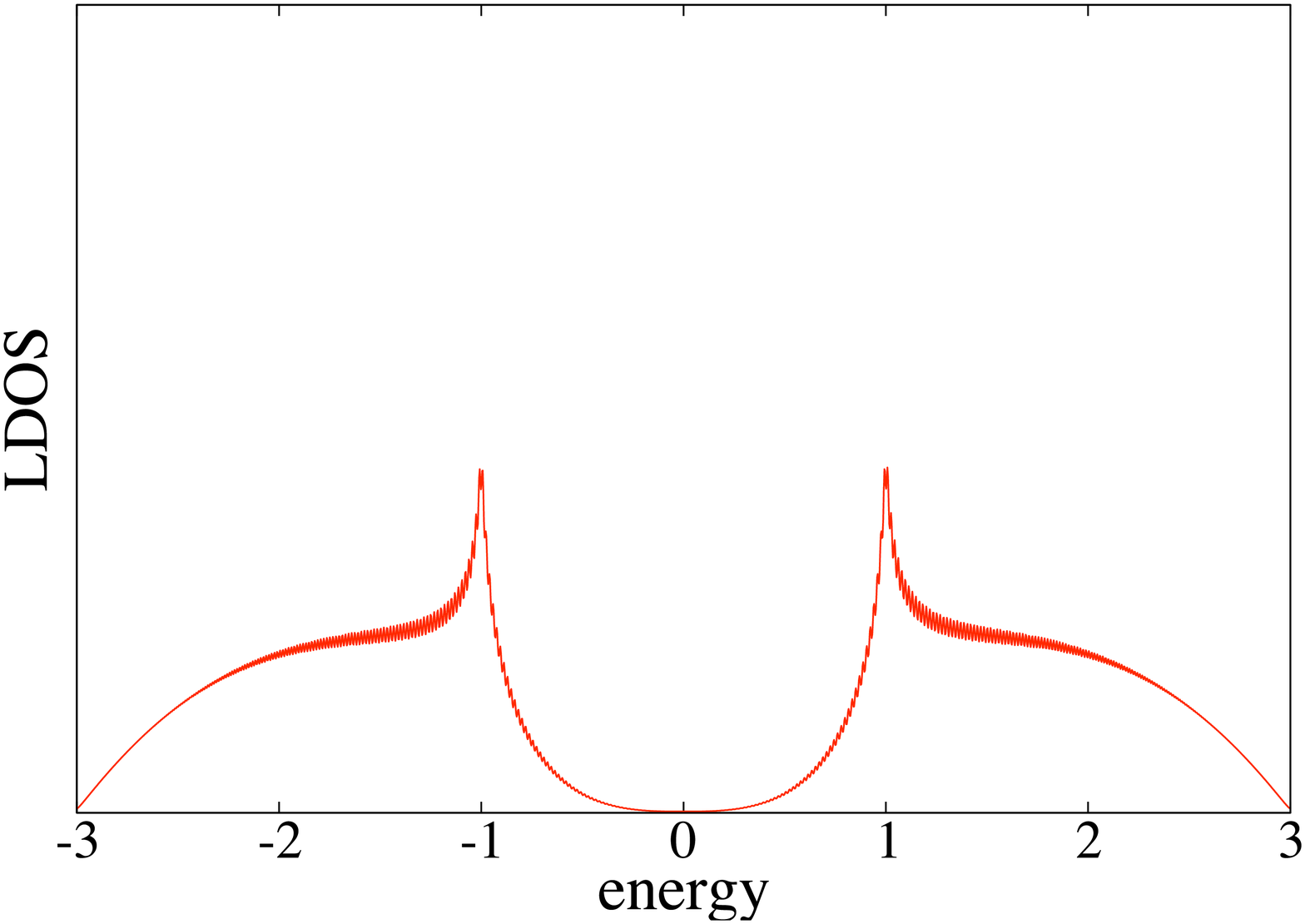}\\
		\end{center}
	\end{minipage}

\vspace{0.5cm}
	\begin{minipage}[t]{0.495\textwidth}
		\begin{center}
			(d)\\
			3\\
			\includegraphics[scale=0.17]
			{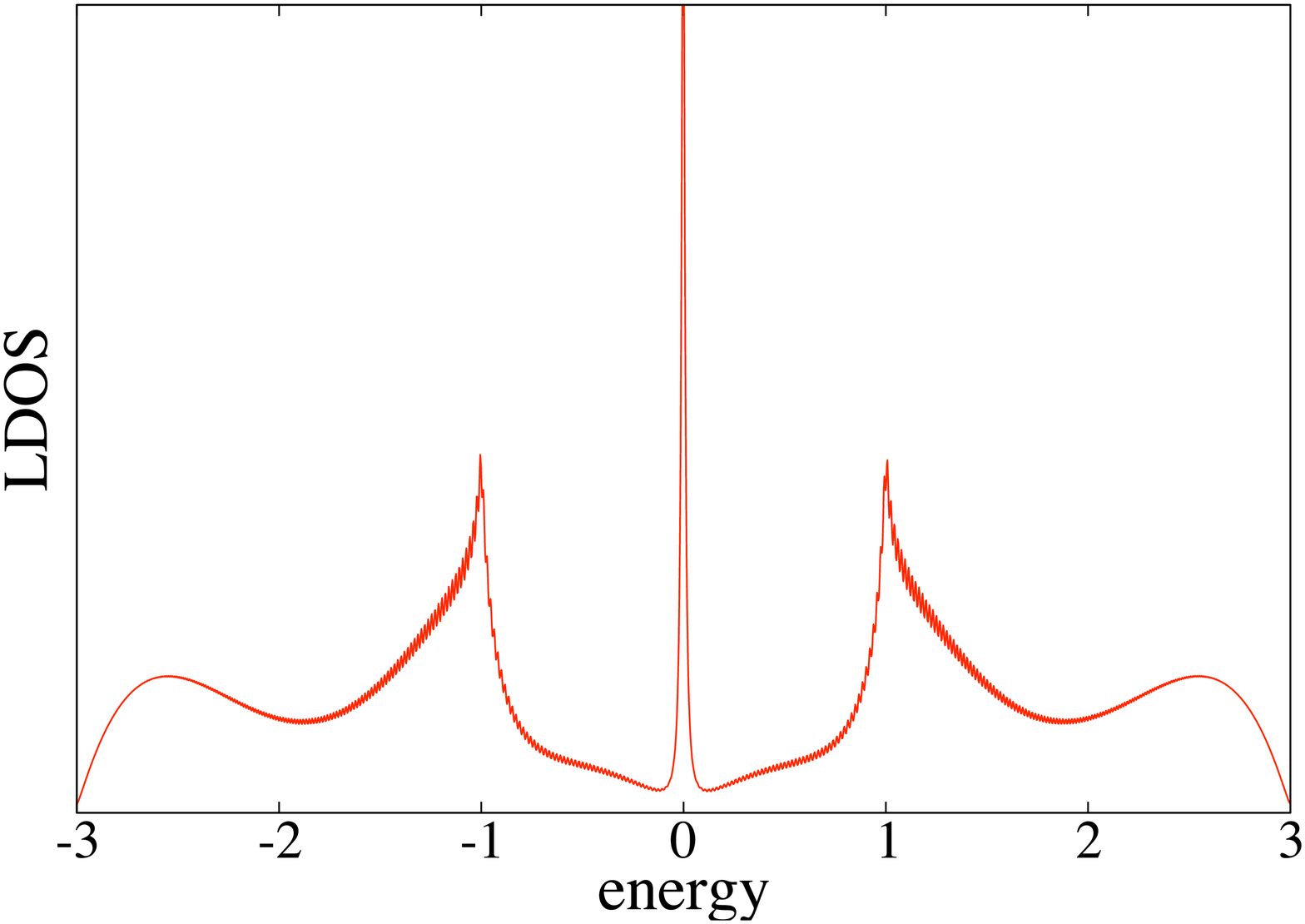}\\
		\end{center}
	\end{minipage}
	\begin{minipage}[t]{0.495\textwidth}
		\begin{center}
			(e)\\
			4\\
			\includegraphics[scale=0.17]
			{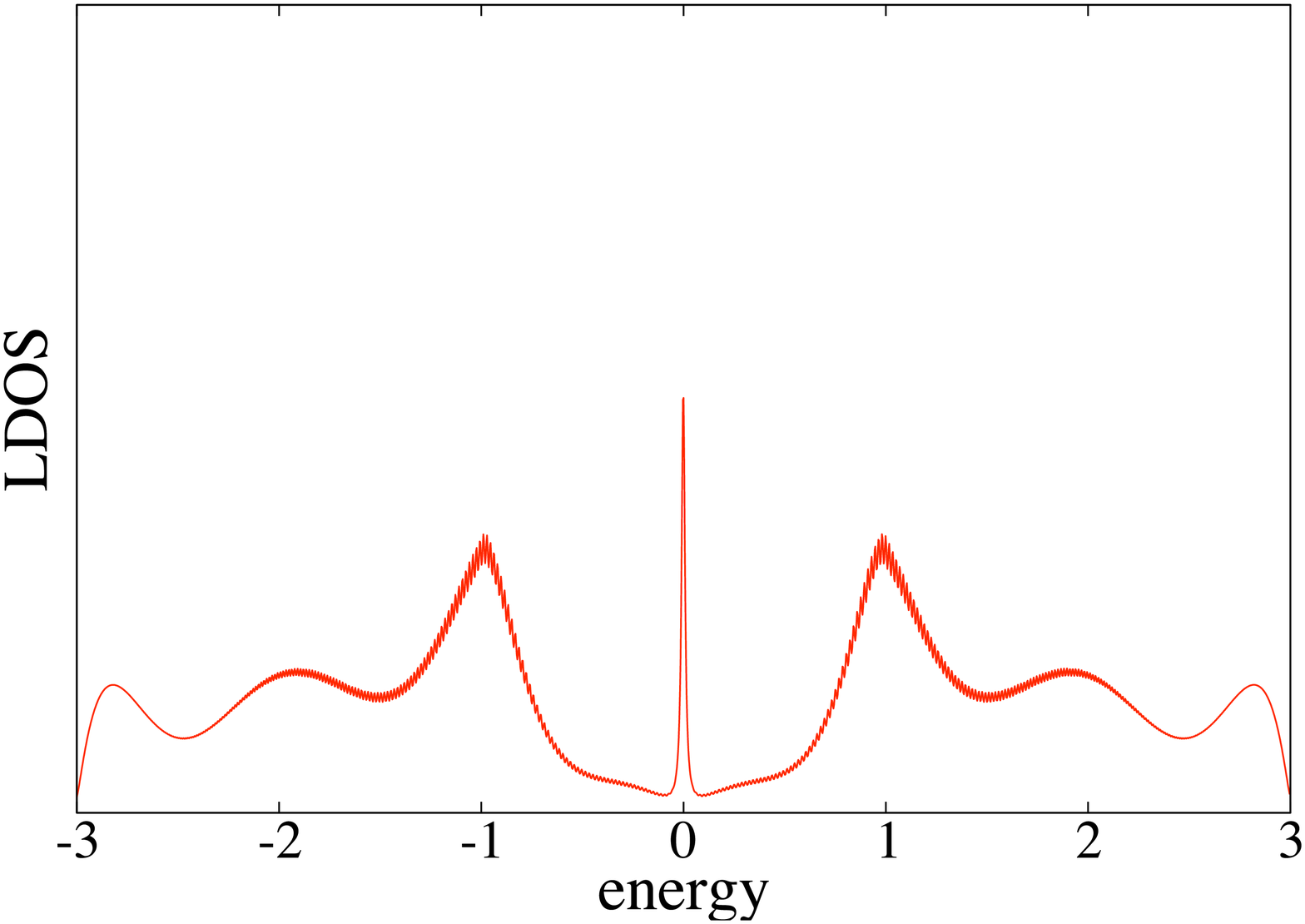}\\
		\end{center}
	\end{minipage}
	\caption{(a) The strcture of a single zz edge.
	A broken line represents a unit cell.
	(b), (c), (d) and (e) display the LDOS
	at the site 1, site 2, site 3, and site 4, respectively.
	The number indicated above each graph represents the site number defined in (a).
	}
	\label{zzldos}
\end{figure}

To confirm the validity of our approach using the recursion method,
we calculate the LDOS in the presence of an ideal single zz or ac edge.
We set $N=1000$, $w/t=300$ and $\delta/t=0.01$ throughout this paper.
We first consider the case with a single zz edge.
The site indices in the unit cell are given in \fref{zzldos}(a).
We display the LDOS at the sites 1, 2, 3, and 4 in \fref{zzldos}(b)-(e).
A peak at $\varepsilon=0$ exists at the site 1 on the zz edge.
The LDOS also possesses a zero-energy peak
at sites on the sublattice which includes the site 1.
We see that the peak decays with increasing the distance from the edge.
At the sites belonging to the other sublattice, such as the site 2,
a peak does not appear at $\varepsilon=0$.
These results are consistent with the presence of edge states at $\varepsilon=0$.
The decay of the zero-energy peak reflects the fact that
an edge state has a finite penetration depth.

\begin{figure}[t]
	\begin{center}
		(a)\\
		\includegraphics[scale=0.6]{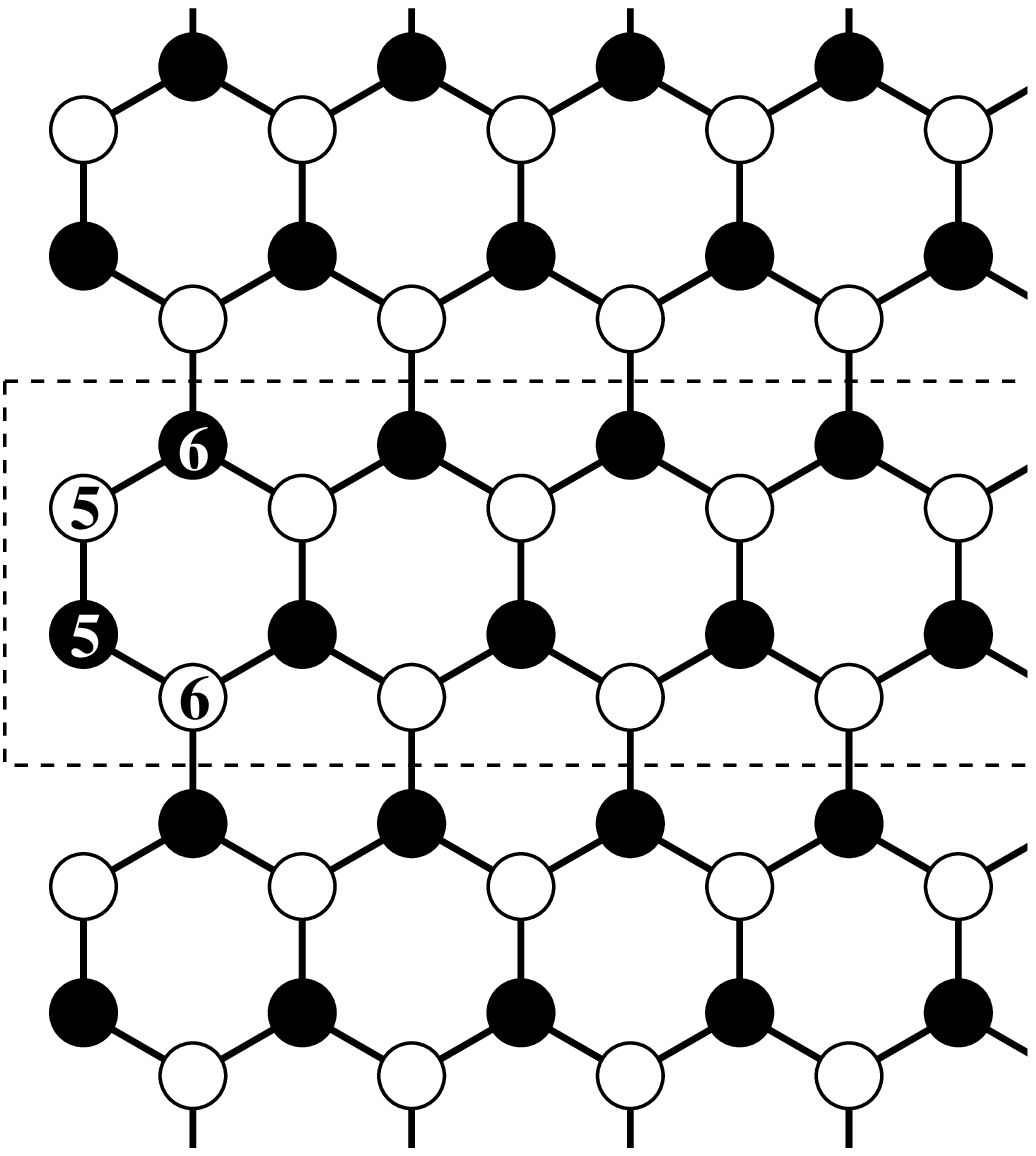}
	\end{center}
	\begin{minipage}[t]{0.495\textwidth}
		\begin{center}
		(b)\\
			5\\
			\includegraphics[scale=0.17]
			{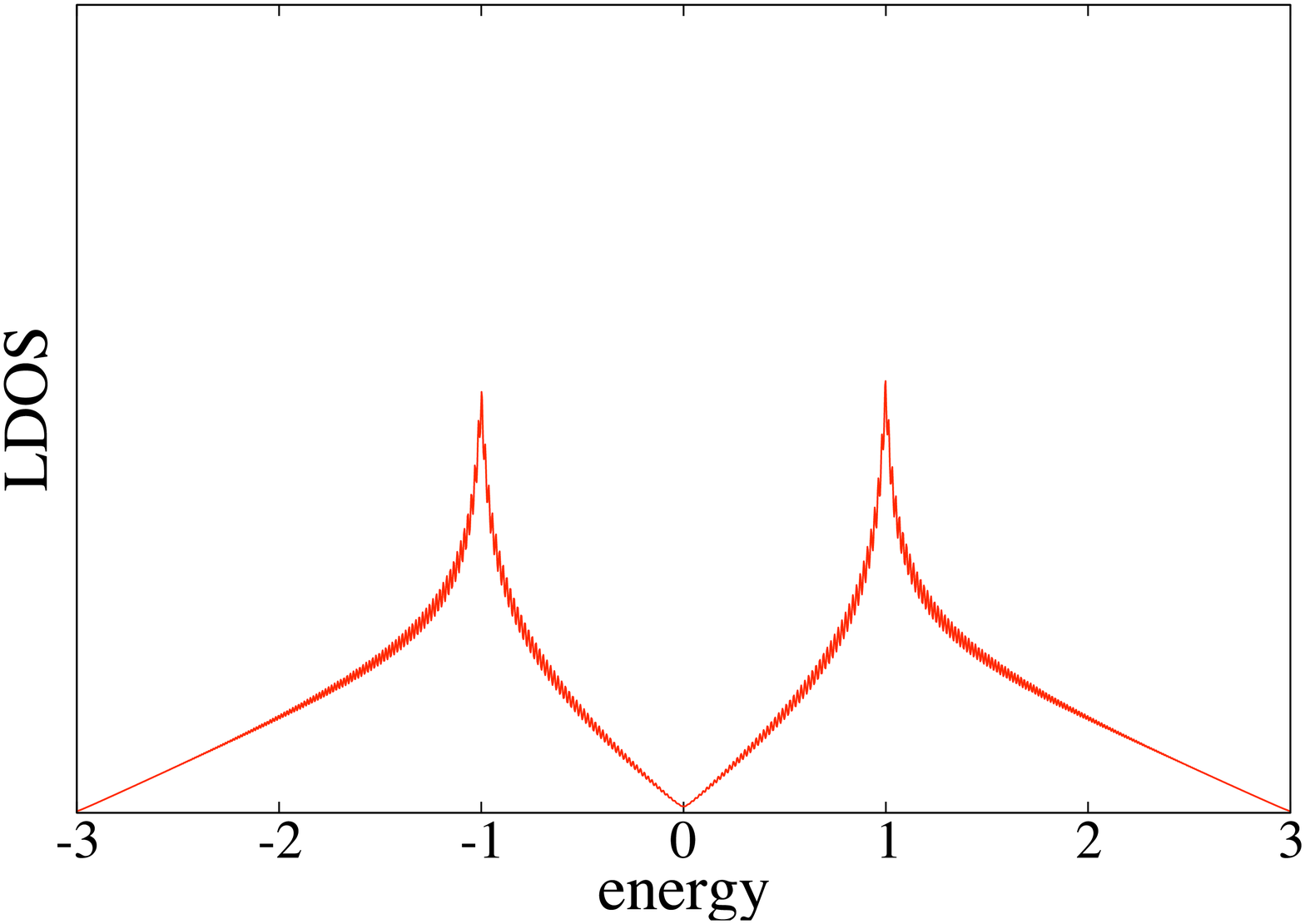}\\
		\end{center}
	\end{minipage}
	\begin{minipage}[t]{0.495\textwidth}
		\begin{center}
		(c)\\
			6\\
			\includegraphics[scale=0.17]
			{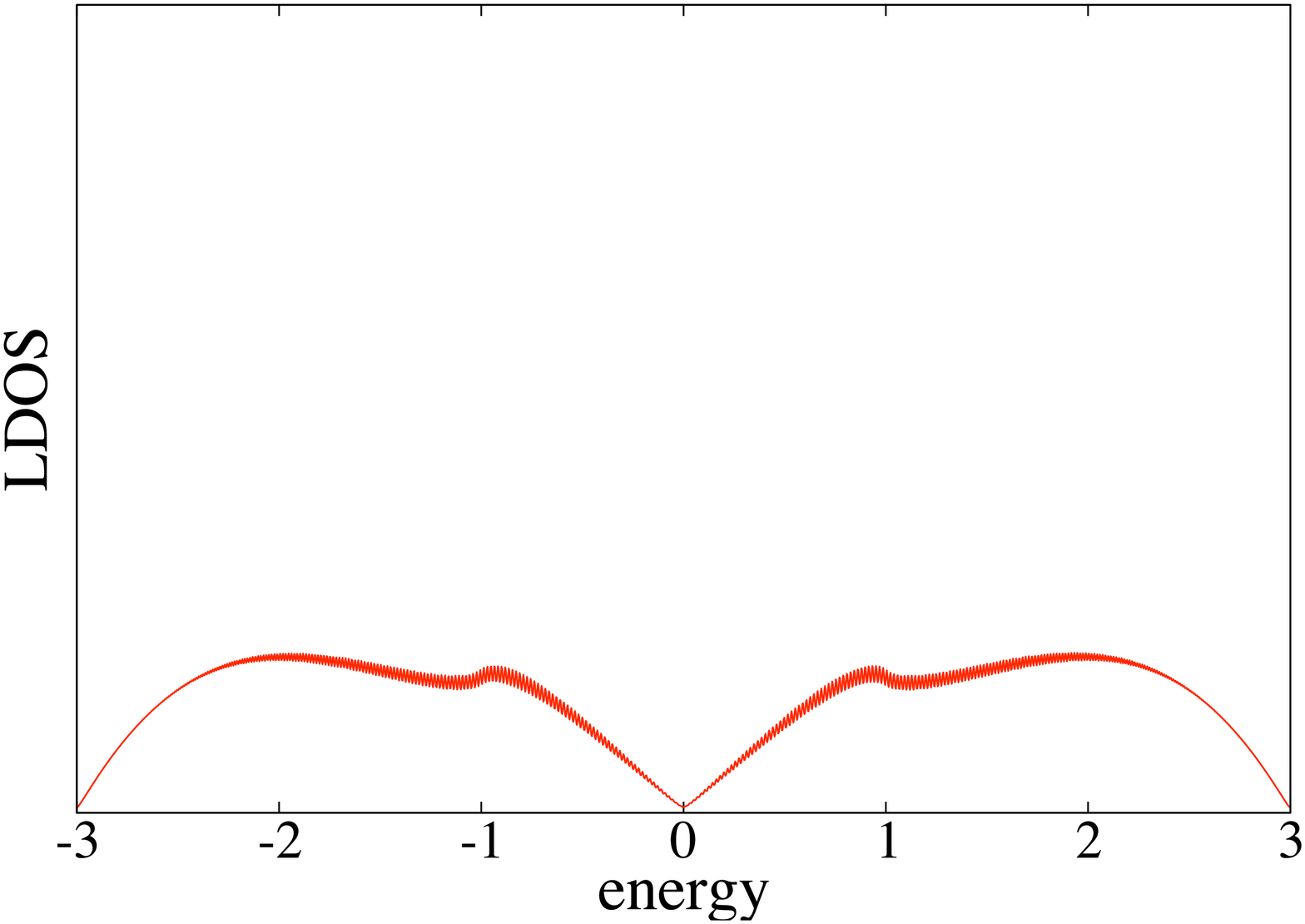}\\
		\end{center}
	\end{minipage}

	\caption{(a) The strcture of a single ac edge.
	A broken line represents a unit cell.
	(b) and (c) display the LDOS at the site 5 and site 6, respectively.
	The number indicated above each graph represents the site number defined in (a).
	}
	\label{acldos}
\end{figure}

We next consider the case with a single ac edge.
Figure \ref{acldos} shows the LDOS in the presence of a single ac edge.
We do not observe a peak of the LDOS at $\varepsilon=0$.
This is consistent with the absence of edge state in the single ac edge case.

\subsection{The LDOS in the presence of a corner edge}
\begin{figure}[t]
	\begin{center}
		(a)\\
		\includegraphics[scale=0.6]{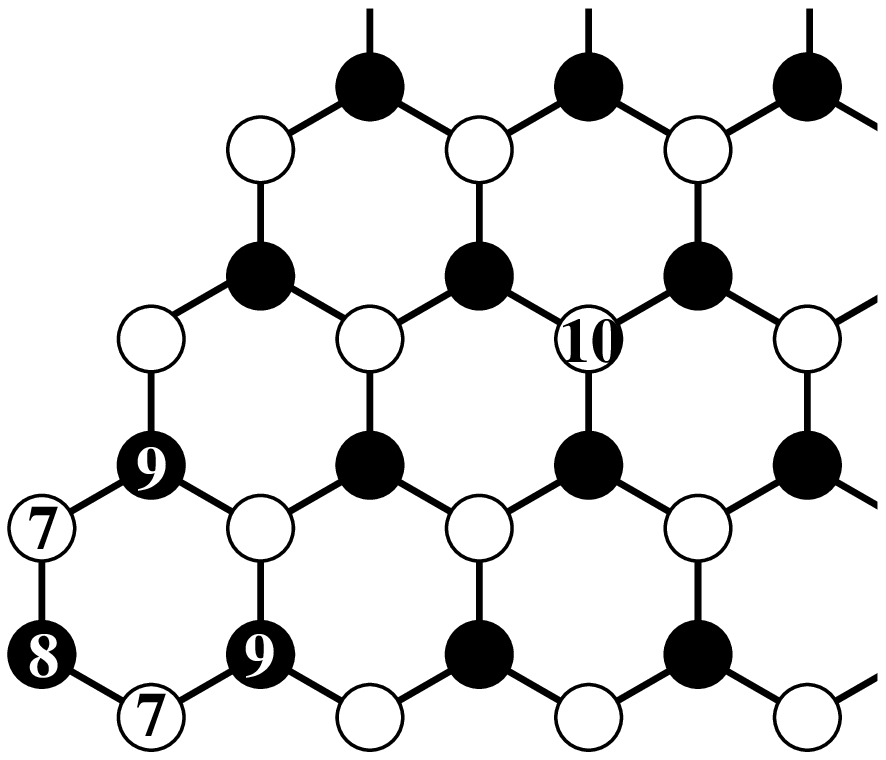}
	\end{center}
	\begin{minipage}[t]{0.495\textwidth}
		\begin{center}
		(b)\\
			7\\
			\includegraphics[scale=0.17]
			{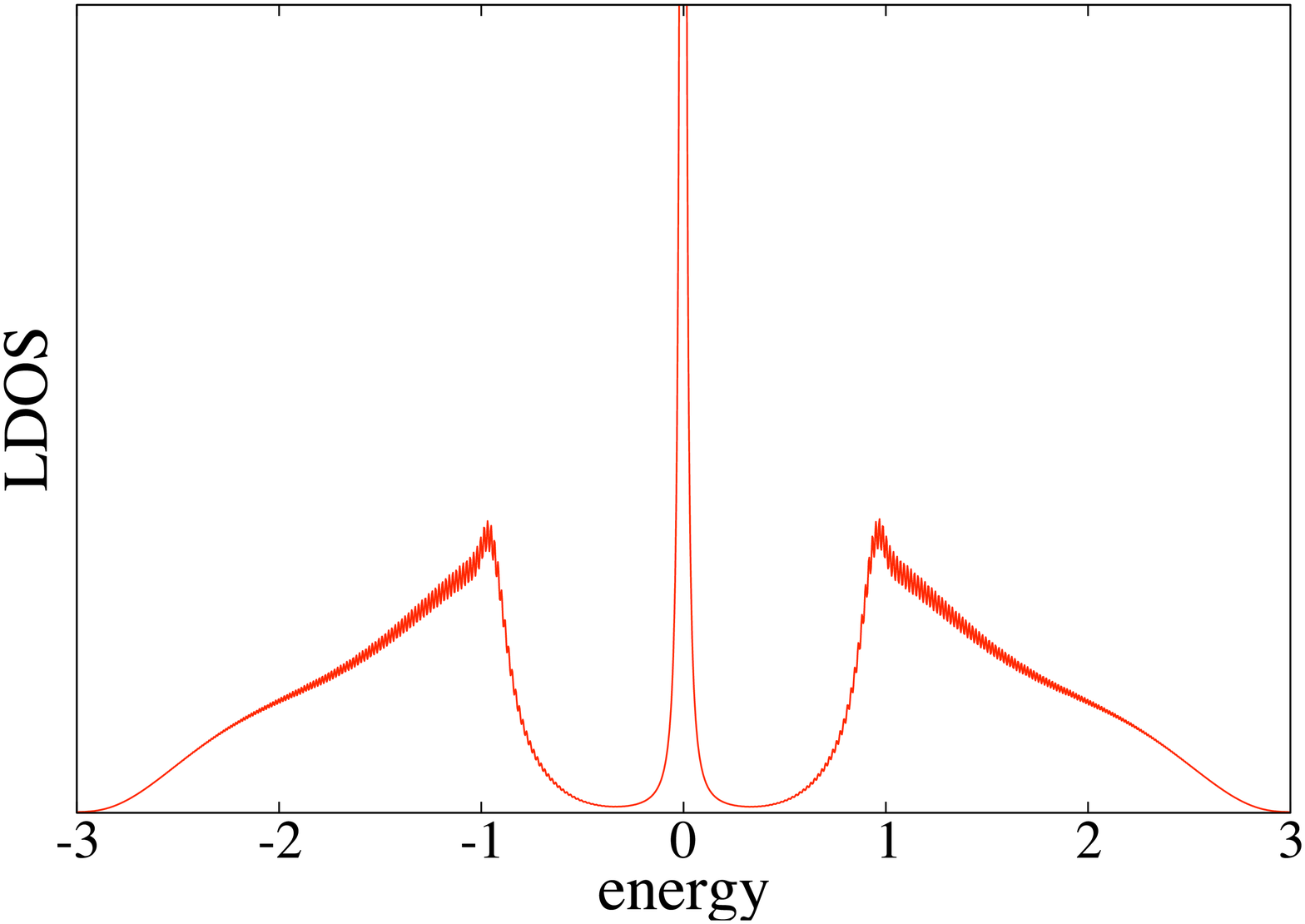}\\
		\end{center}
	\end{minipage}
	\begin{minipage}[t]{0.495\textwidth}
		\begin{center}
		(c)\\
			8\\
			\includegraphics[scale=0.17]
			{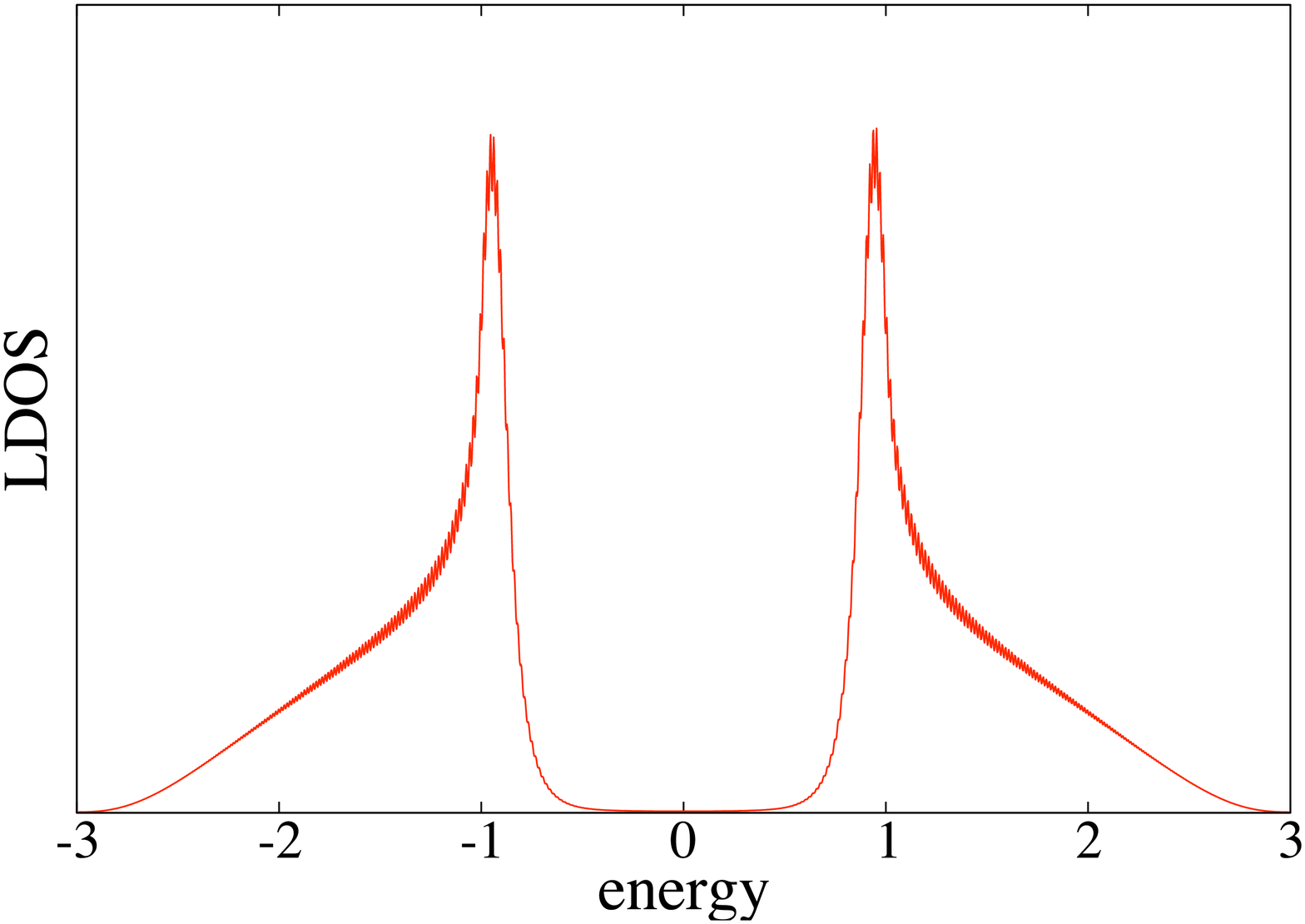}\\
		\end{center}
	\end{minipage}

\vspace{0.5cm}
	\begin{minipage}[t]{0.495\textwidth}
		\begin{center}
		(d)\\
			9\\
			\includegraphics[scale=0.17]
			{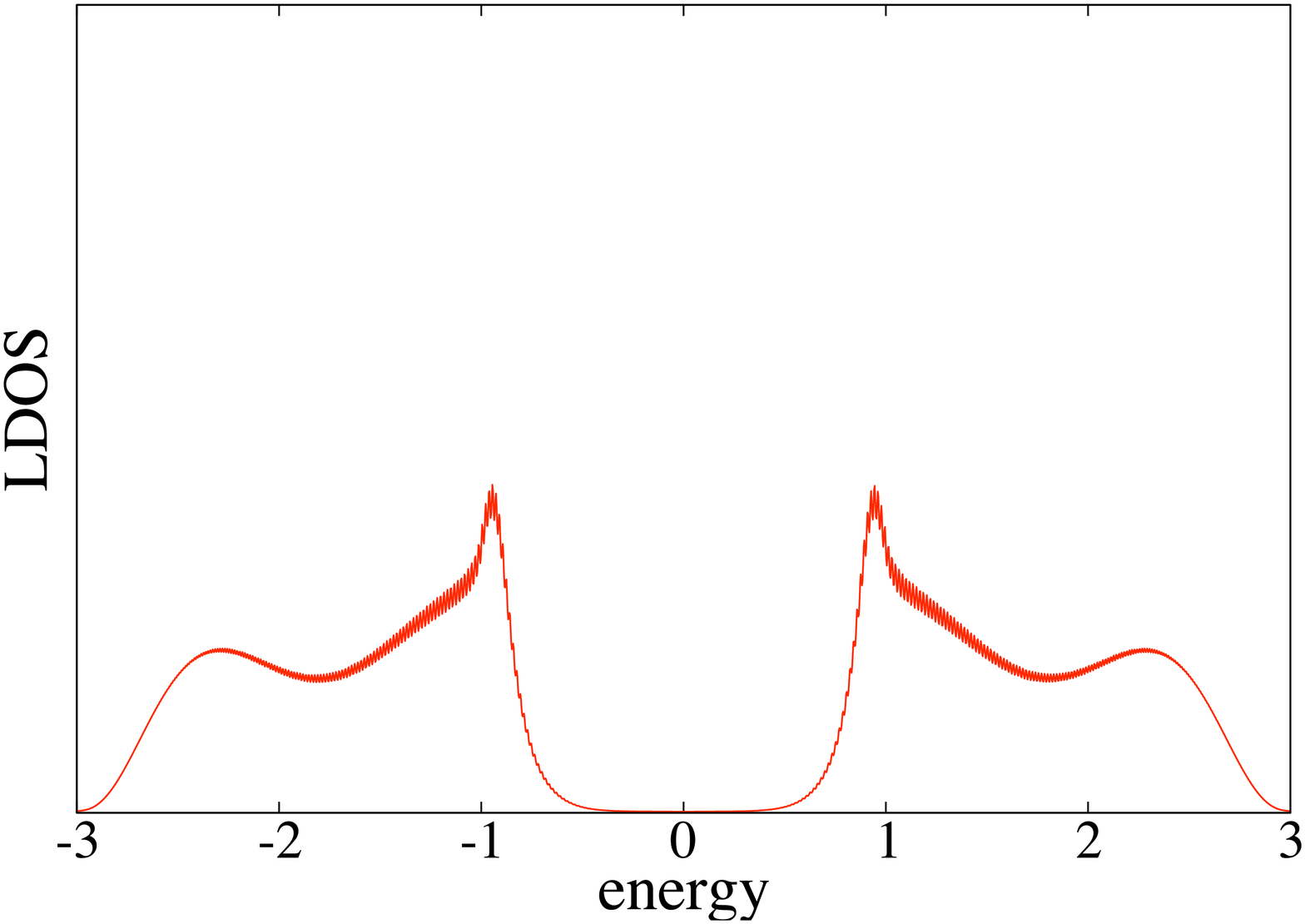}\\
		\end{center}
	\end{minipage}
	\begin{minipage}[t]{0.495\textwidth}
		\begin{center}
		(e)\\
			10\\
			\includegraphics[scale=0.17]
			{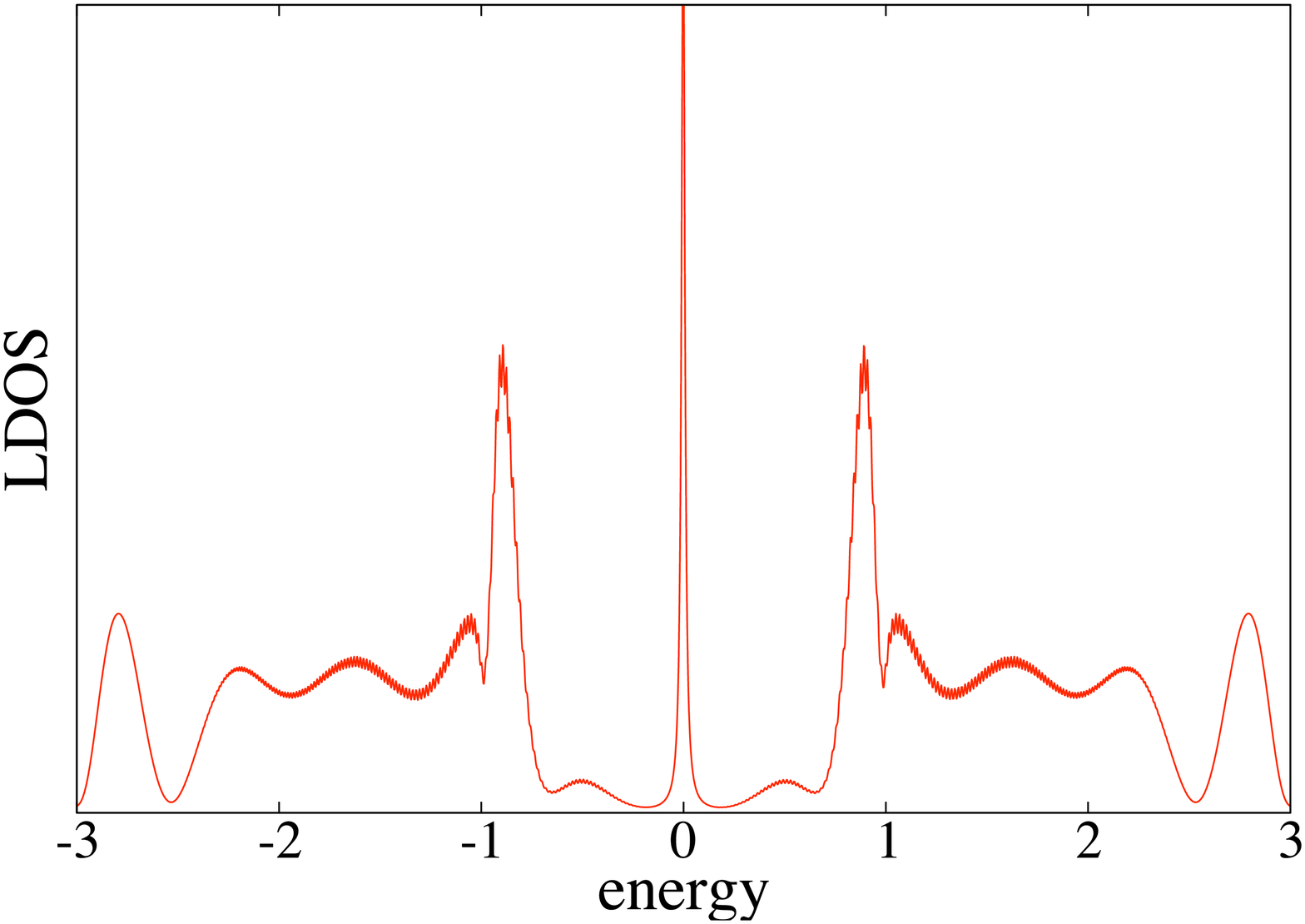}\\
		\end{center}
	\end{minipage}
	\caption{(a) The strcture of the 60$^\circ$ corner edge.
	(b), (c), (d) and (e) display the LDOS at the site 7, site 8, site 9, and site 10, respectively.
	The number indicated above each graph represents the site number defined in (a).
	}
	\label{60ldos}
\end{figure}

\noindent
\textit{{\rm (i)} 60$^\circ$ corner edge.-}
Figure \ref{60ldos} shows the LDOS at several sites in the presence of the 60$^\circ$ corner edge consisting of two zz edges.
From this figure, we can see the appearance of edge states at $\varepsilon=0$.
As shown in \fref{60ldos}(b) and (e), 
a zero-energy peak exists at the sites 7 and 10 belonging to a same sublattice.
Let us compare the LDOS at the site 10 (\fref{60ldos}(e)) with 
that at the site 4 in the single zz edge case (\fref{zzldos}(e)).
Note that the distance from the zz edge to the site of our interest is equivalent in both the cases.
We observe that 
the peak of the LDOS at the site 10
is higher than that at the site 4 in the single zz edge case.
We consider that this enhancement of the zero-energy peak at the site 10 
is caused by a superposition of edge states 
at one zz edge and those at the other edge, i.e. constructive
interference between two edge states.

\begin{figure}[t]
	\begin{center}
		(a)\\
		\includegraphics[scale=0.6]{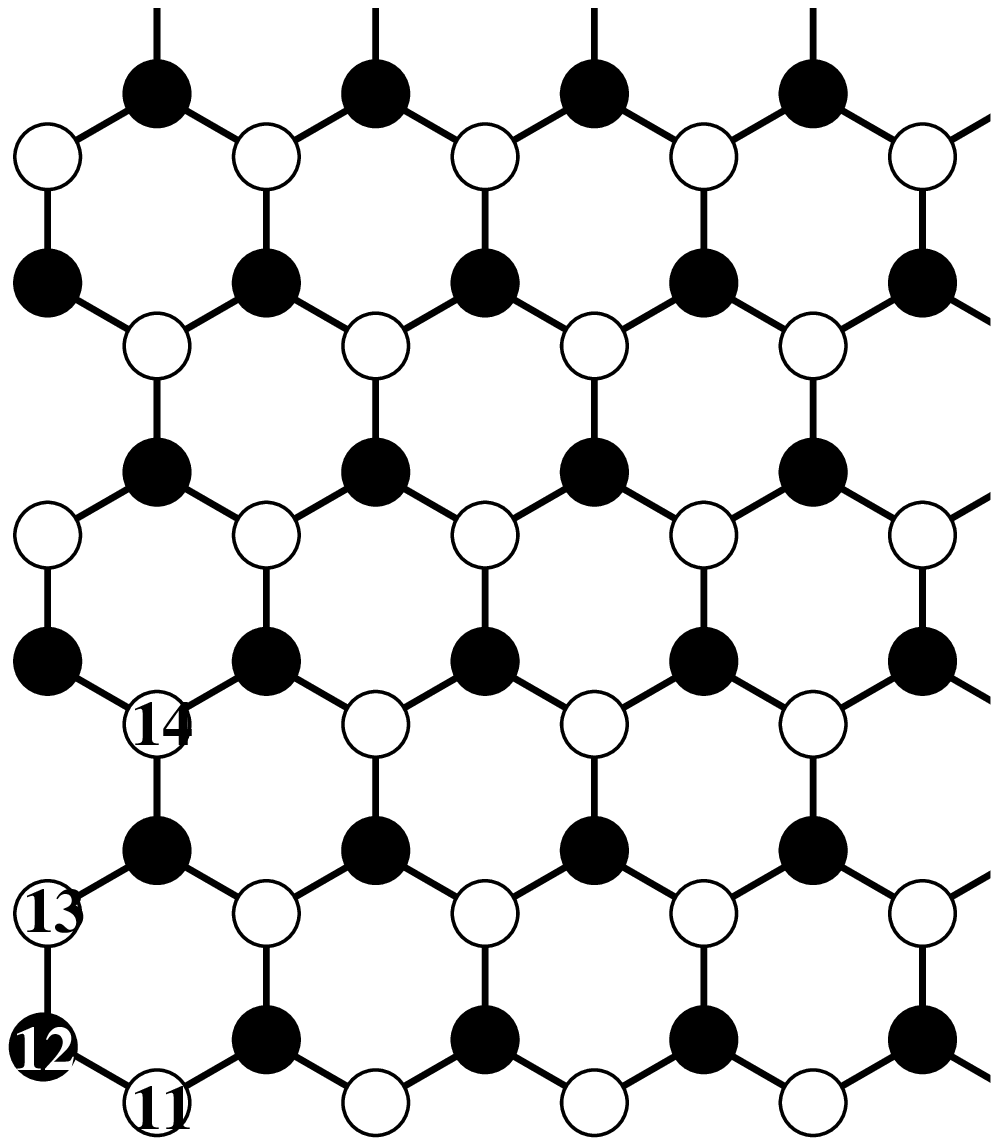}
	\end{center}
	\begin{minipage}[t]{0.495\textwidth}
		\begin{center}
			(b)\\
			11\\
			\includegraphics[scale=0.17]
			{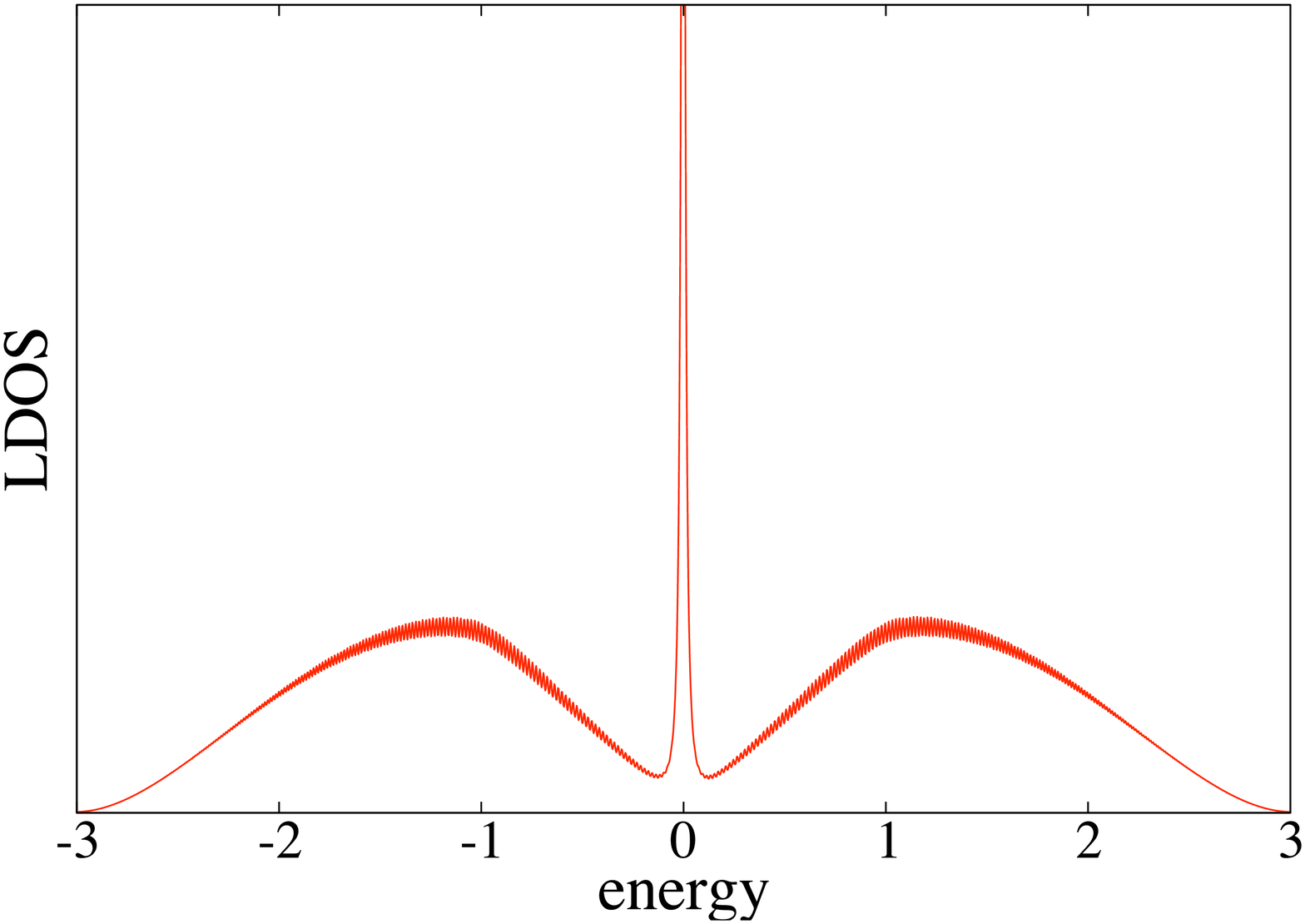}\\
		\end{center}
	\end{minipage}
	\begin{minipage}[t]{0.495\textwidth}
		\begin{center}
			(c)\\
			12\\
			\includegraphics[scale=0.17]
			{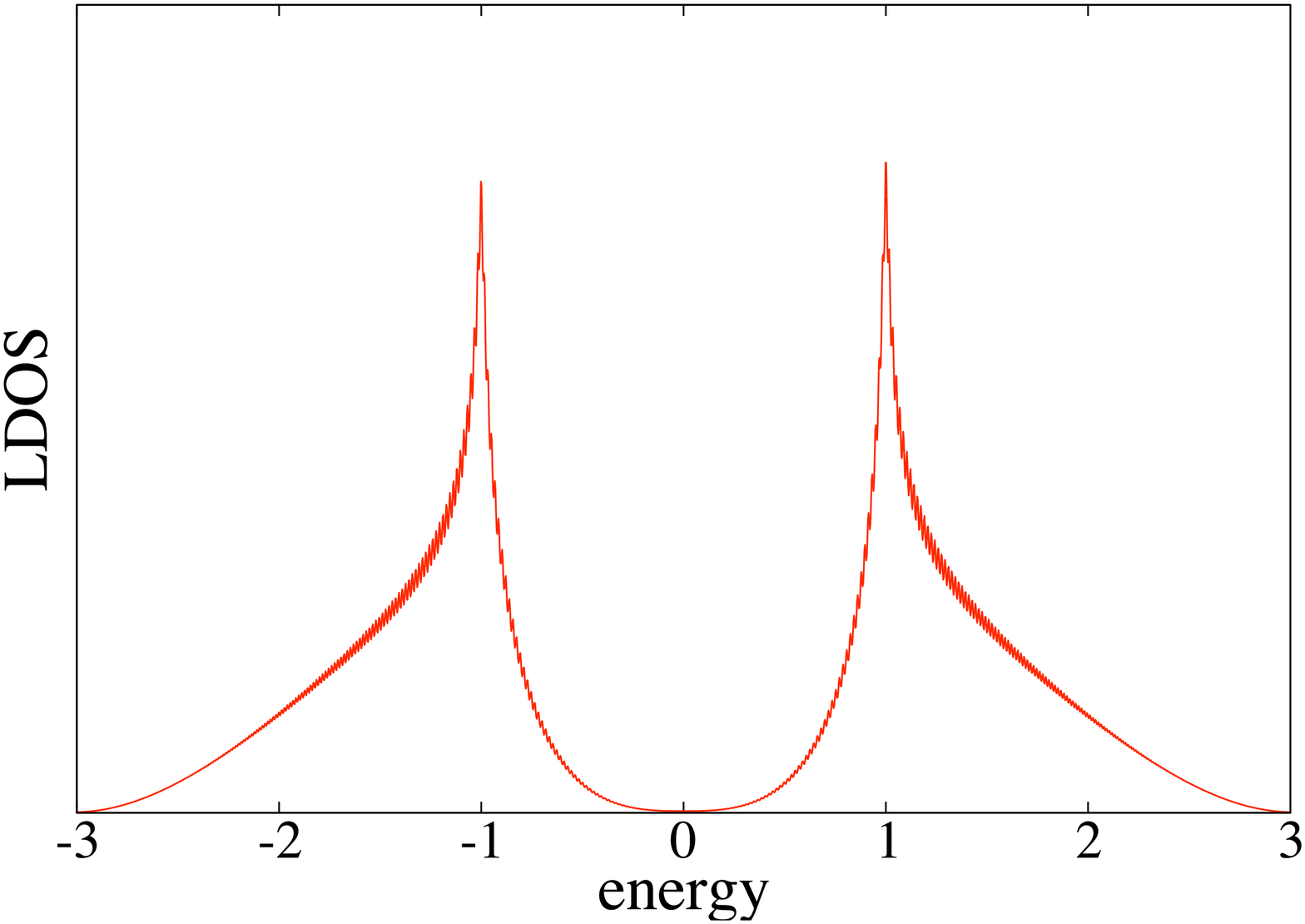}\\
		\end{center}
	\end{minipage}

\vspace{0.5cm}
	\begin{minipage}[t]{0.495\textwidth}
		\begin{center}
			(d)\\
			13\\
			\includegraphics[scale=0.17]
			{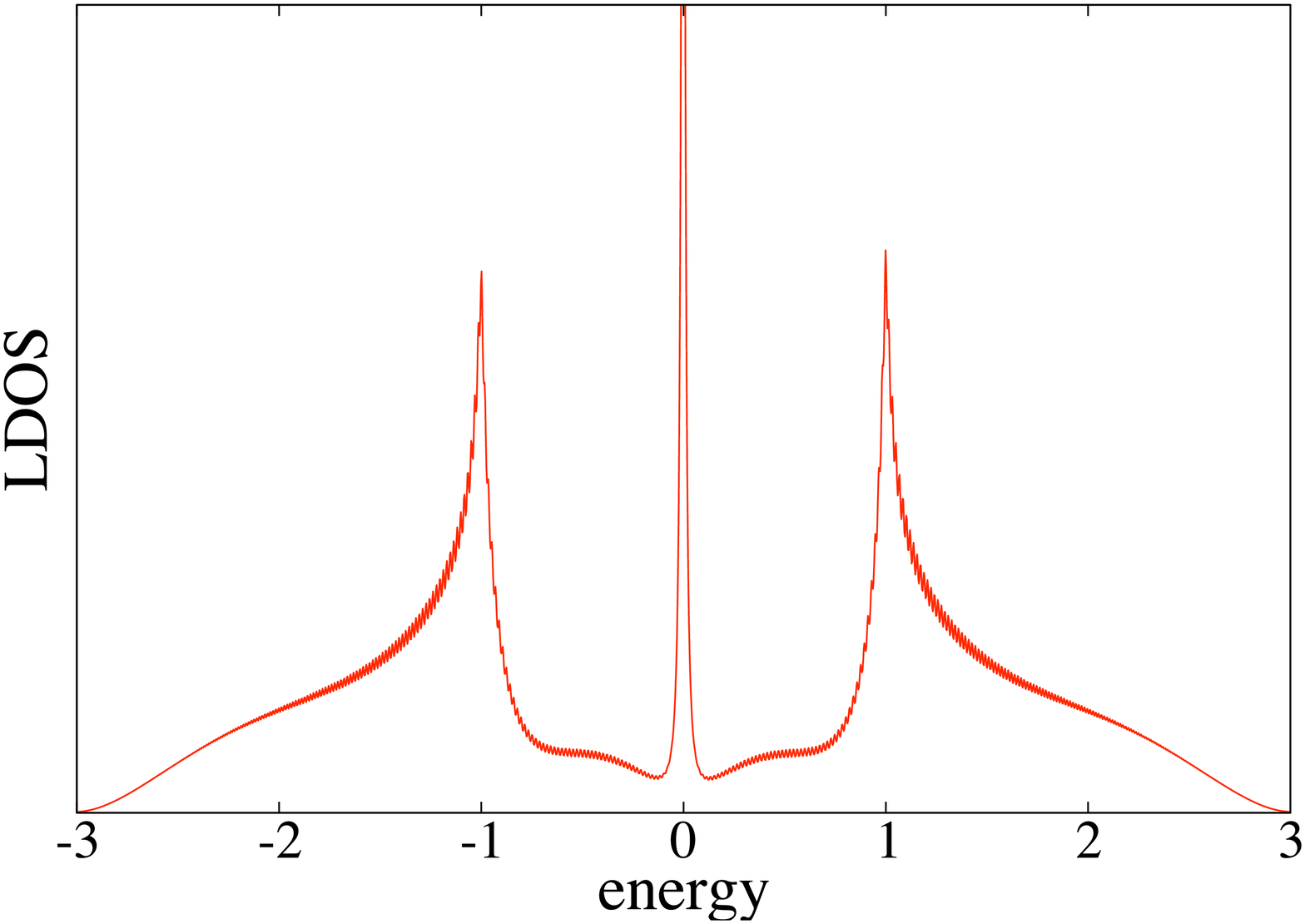}\\
		\end{center}
	\end{minipage}
	\begin{minipage}[t]{0.495\textwidth}
		\begin{center}
			(e)\\
			14\\	
			\includegraphics[scale=0.17]
			{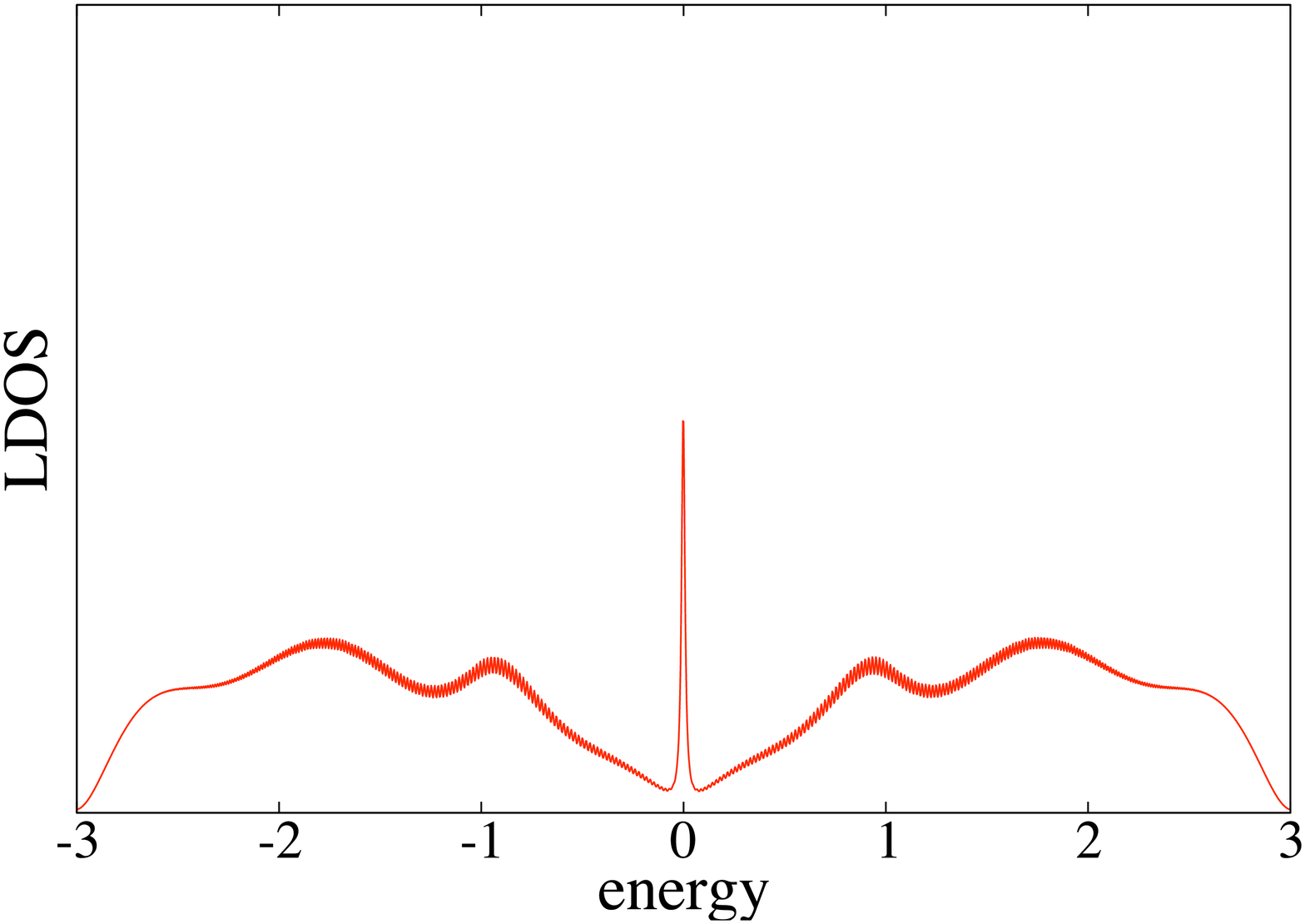}\\
		\end{center}
	\end{minipage}
	\caption{(a) The strcture of the 90$^\circ$ corner edge.
	(b), (c), (d) and (e) display the LDOS at the site 11, site 12, site 13, and site 14, respectively.
	The number indicated above each graph represents the site number defined in (a).
	}
	\label{90ldos}
\end{figure}

\noindent
\textit{{\rm (ii)} 90$^\circ$ corner edge.-}
Figure \ref{90ldos} shows the LDOS at several sites in the presence of the 90$^\circ$ corner edge.
From this figure,
we see that edge states appear at $\varepsilon=0$.
As shown in \fref{90ldos}(b), (d), and (e), 
a zero-energy peak exists at the sites 11, 13 and 14 belonging to a same sublattice.
Thus, the LDOS near the corner possesses both the character of the LDOS in the single zz edge case and that in the single ac edge case.
Let us focus on the LDOS at the site 13 for example.
The site 13 corresponds to the site 3 in the zz edge case (\fref{zzldos}(d)) 
and the site 5 in the single ac edge case (\fref{acldos}(b)).
Roughly speaking, we can regard that
the LDOS at the site 13 (\fref{90ldos}(d)) is a mixture of the LDOS
at the site 3 (\fref{zzldos}(d)) 
and the site 5 (\fref{acldos}(b)).
The nature similar to this is also observed at other sites.
There is no enhancement of the zero-energy peak of LDOS in contrast to the 60$^\circ$ case.

\begin{figure}[t]
	\begin{center}
		(a)\\
		\includegraphics[scale=0.6]{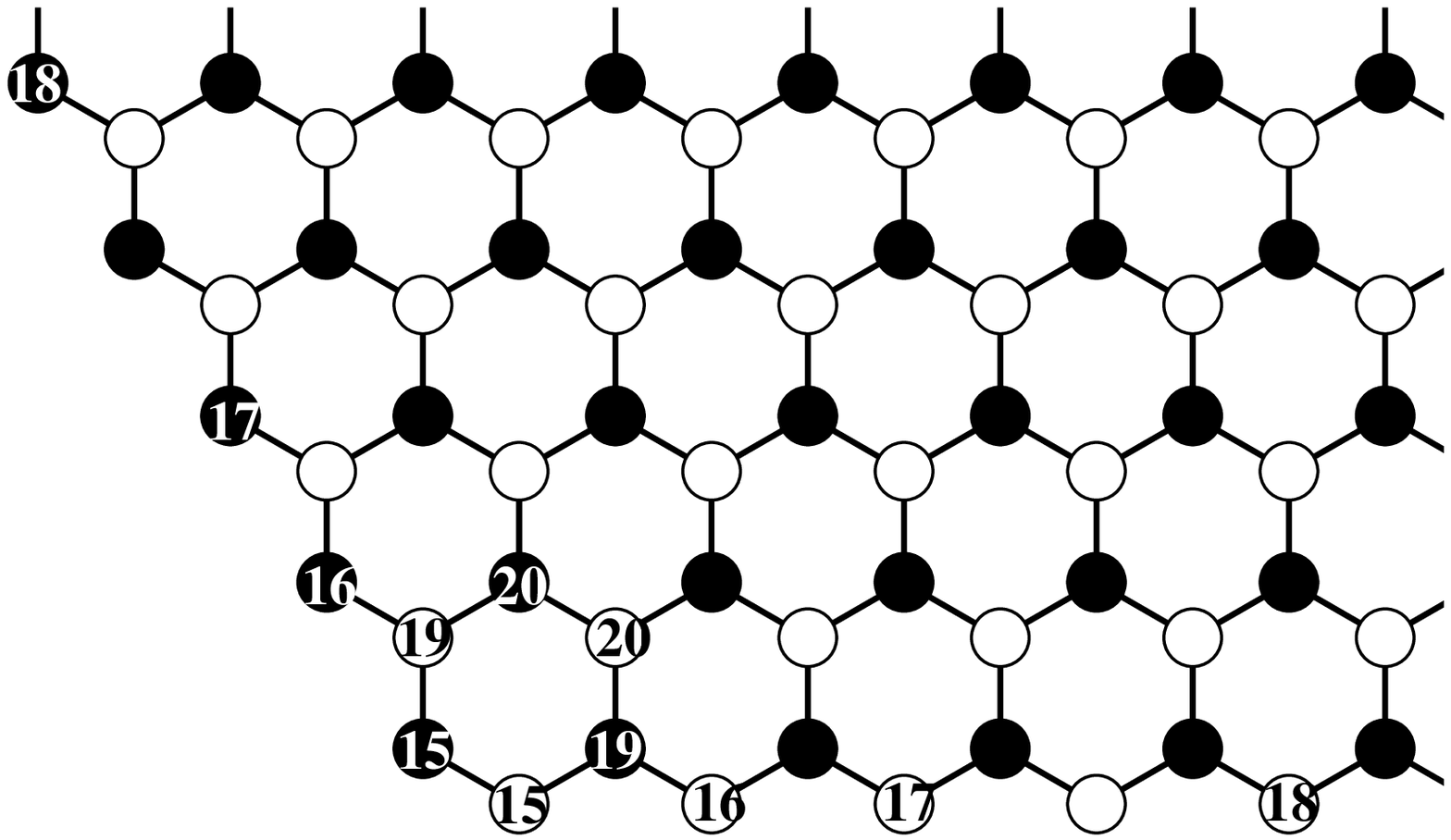}
	\end{center}
	\begin{minipage}[t]{0.495\textwidth}
		\begin{center}
			(b)\\
			15\\
			\includegraphics[scale=0.17]
			{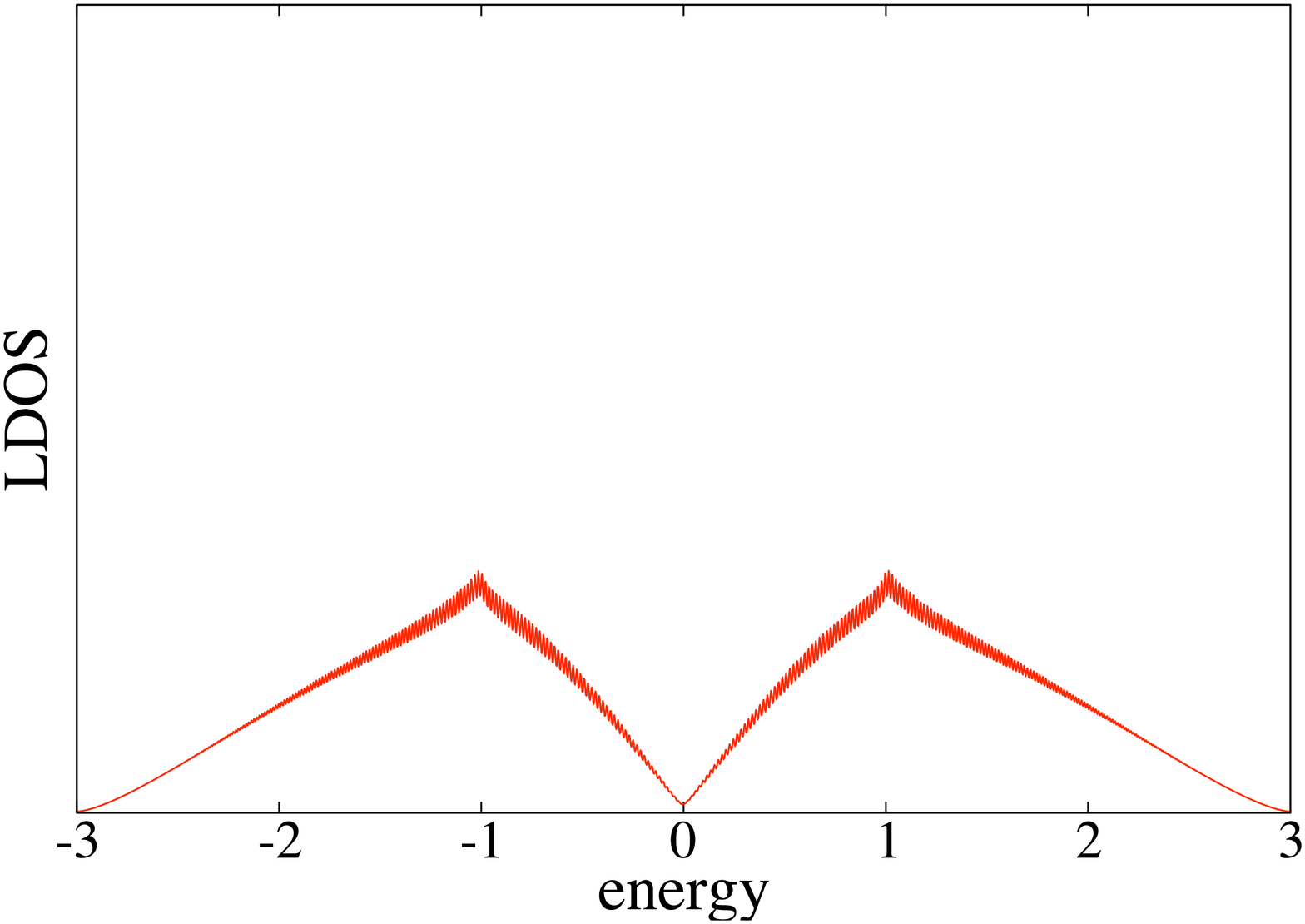}\\
		\end{center}
	\end{minipage}
	\begin{minipage}[t]{0.495\textwidth}
		\begin{center}
			(c)\\
			16\\
			\includegraphics[scale=0.17]
			{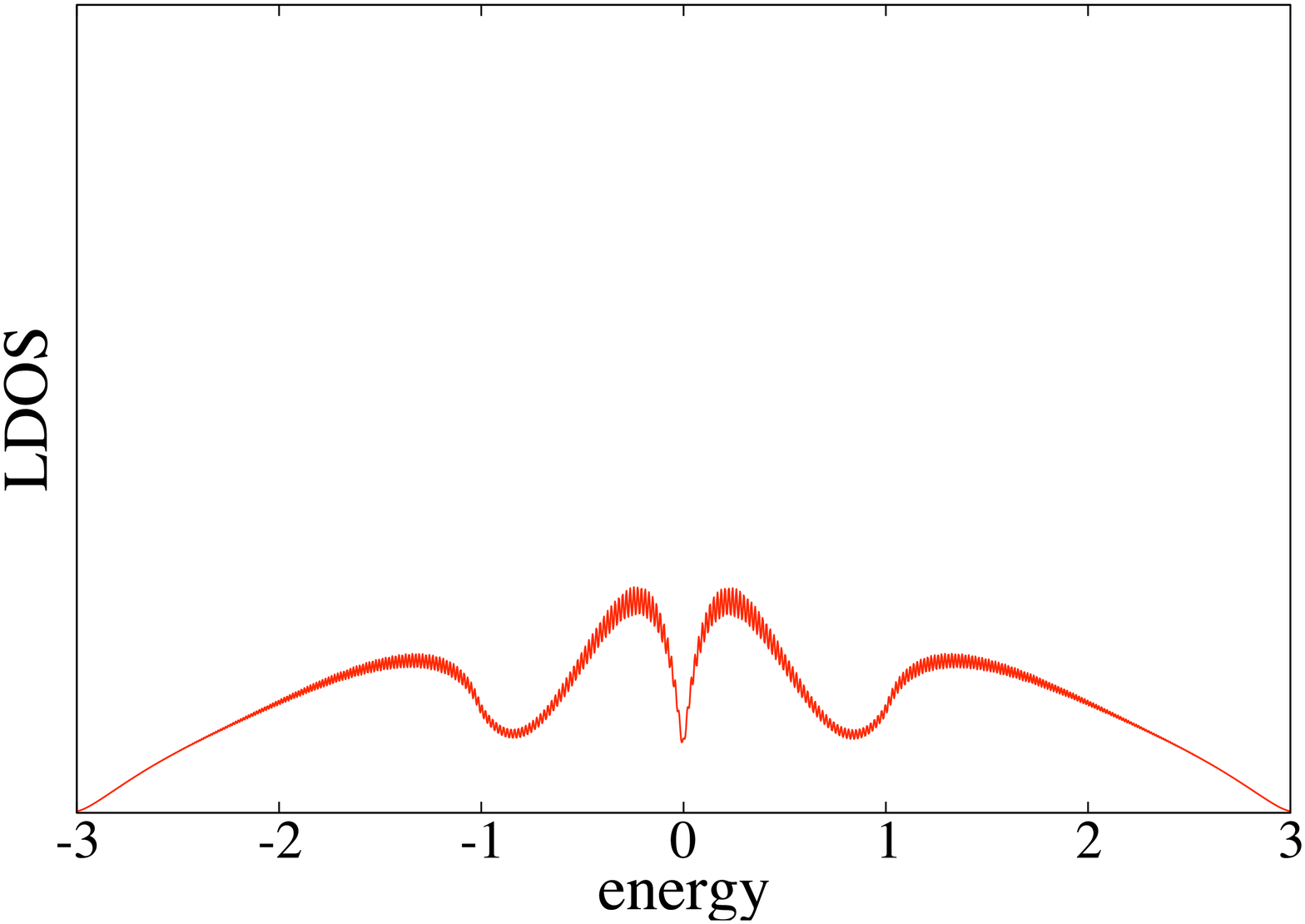}\\
		\end{center}
	\end{minipage}

\vspace{0.3cm}
	\begin{minipage}[t]{0.495\textwidth}
		\begin{center}
			(d)\\
			17\\
			\includegraphics[scale=0.17]
			{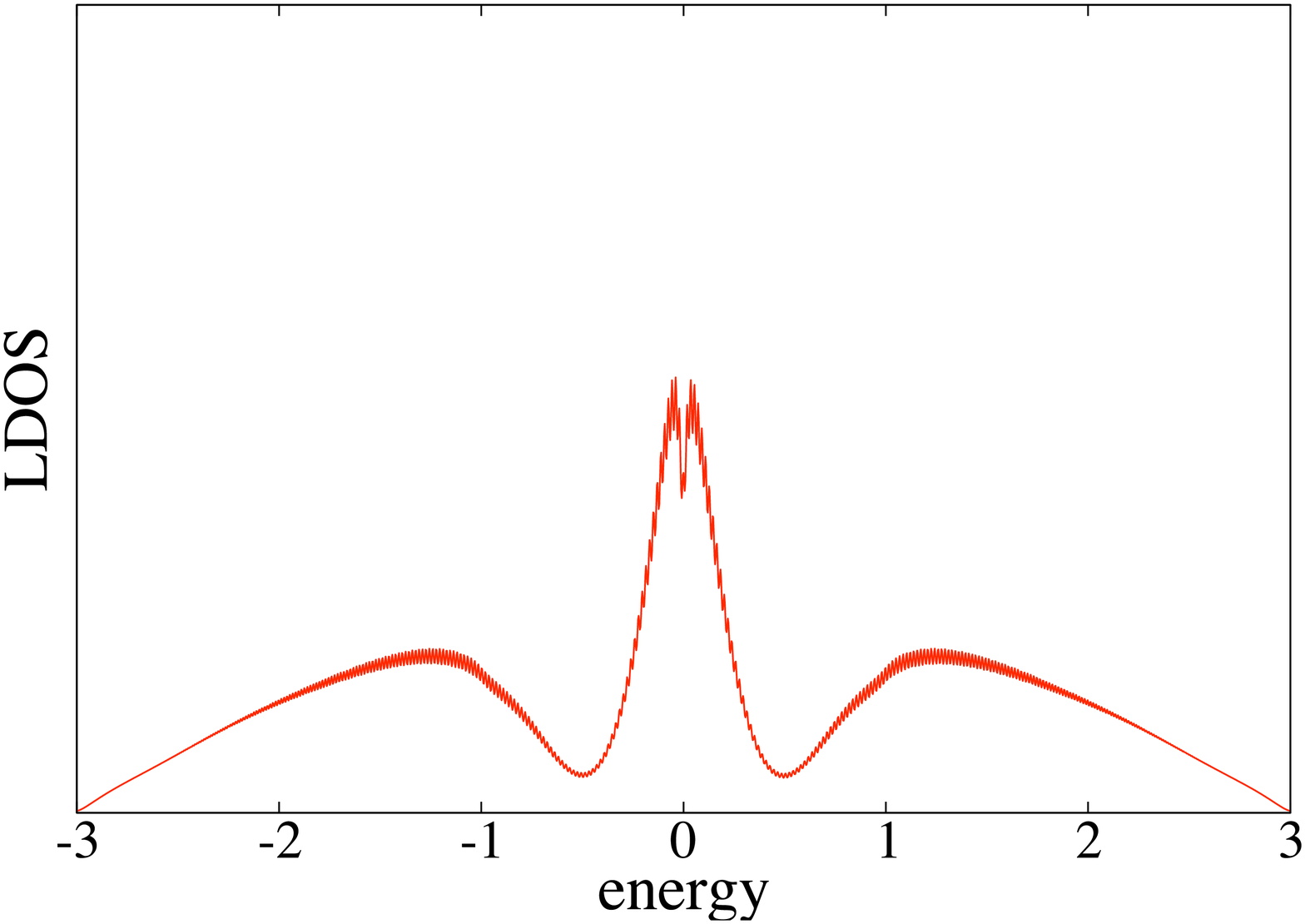}\\
		\end{center}
	\end{minipage}
	\begin{minipage}[t]{0.495\textwidth}
		\begin{center}
			(e)\\
			18\\
			\includegraphics[scale=0.17]
			{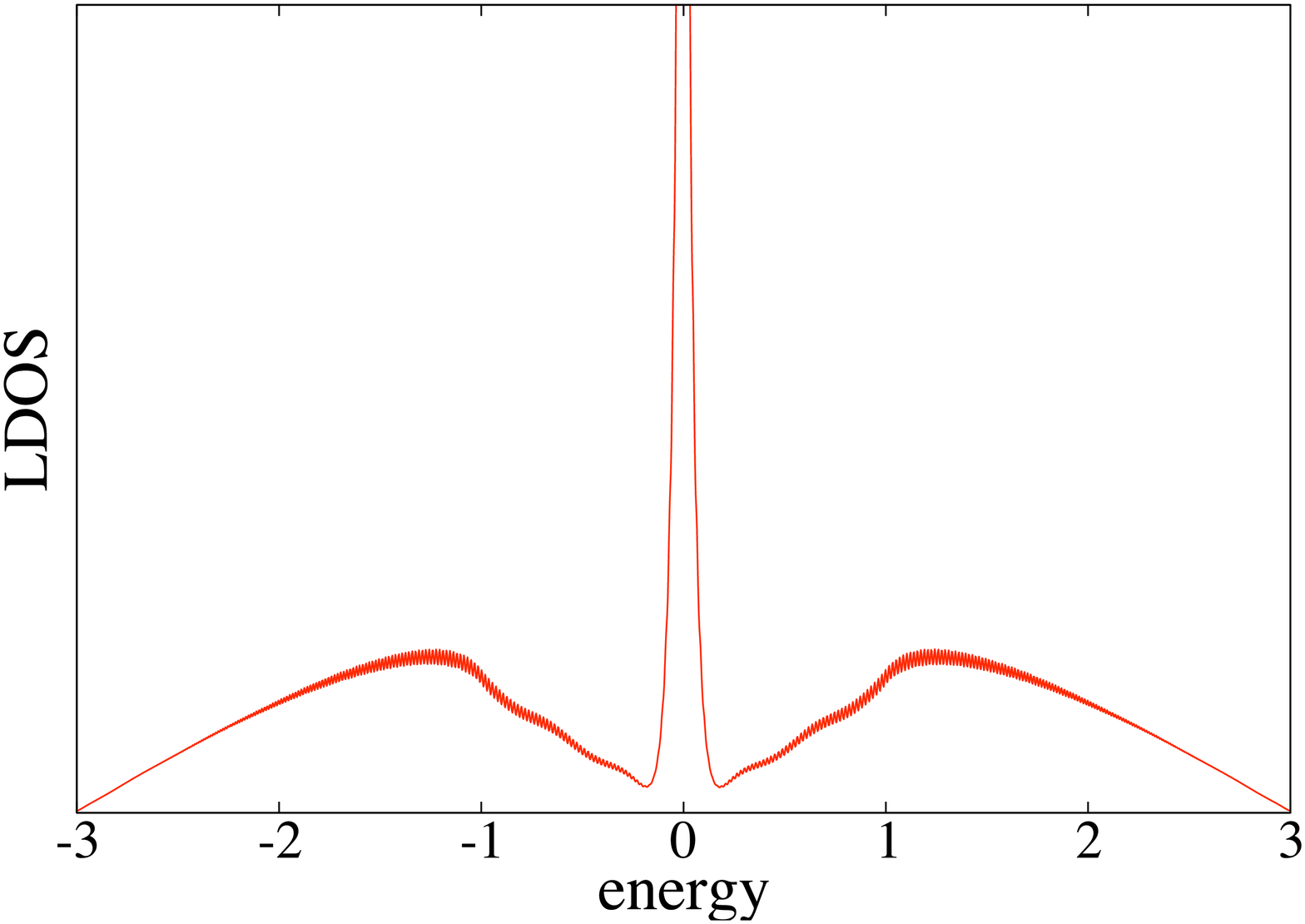}\\
		\end{center}
	\end{minipage}

\vspace{0.3cm}
	\begin{minipage}[t]{0.495\textwidth}
		\begin{center}
			(f)\\
			19\\
			\includegraphics[scale=0.17]
			{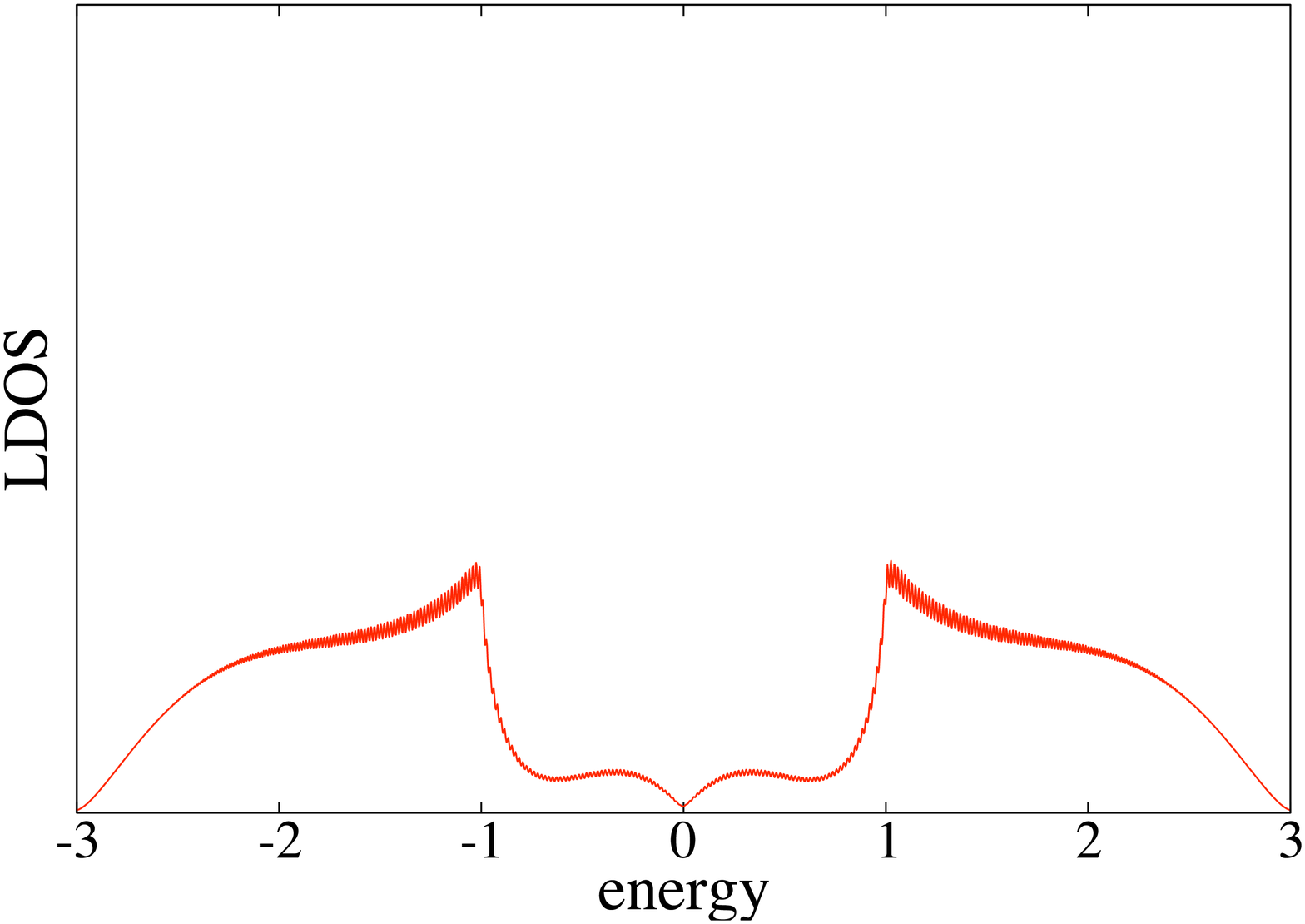}\\
		\end{center}
	\end{minipage}
	\begin{minipage}[t]{0.495\textwidth}
		\begin{center}
			(g)\\
			20\\
			\includegraphics[scale=0.17]
			{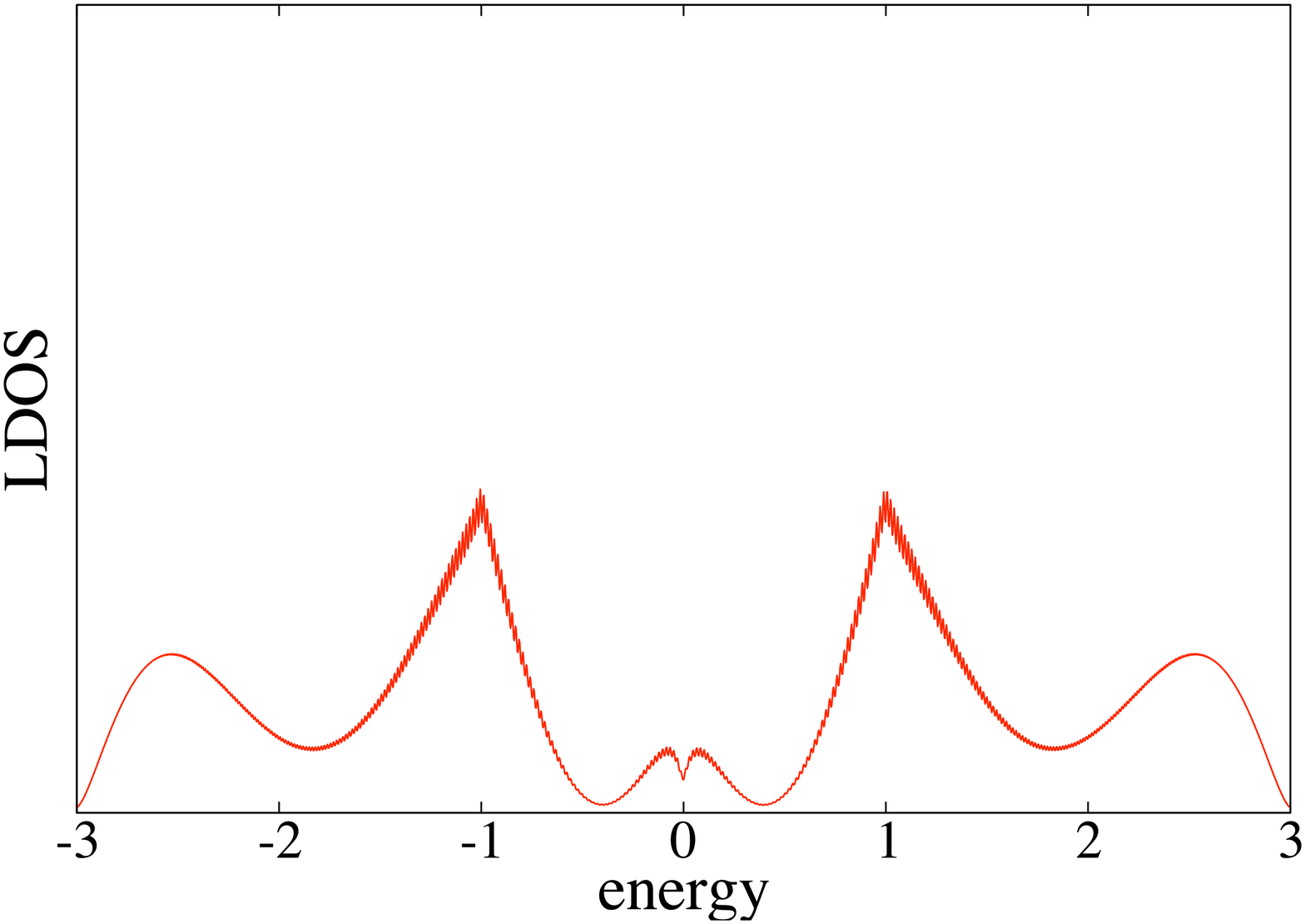}\\
		\end{center}
	\end{minipage}
	\caption{(a) The strcture of the 120$^\circ$ corner edge.
	(b), (c), (d), (e), (f), and (g) display the LDOS at the site 15, site 16, site 17, site 18, site 19, and site 20, respectively.
	The number indicated above each graph represents the site number defined in (a).
	}
	\label{120ldos}
\end{figure}

\noindent
\textit{{\rm (iii)} 120$^\circ$ corner edge.-}
Figure \ref{120ldos} shows the LDOS at several sites in the presence of the 120$^\circ$ corner consisting of two zz edges.
In this case,
peculiar features arise.
The LDOS at the site 15 (\fref{120ldos}(b)) is quite different from that of the site 1 in the single zz edge case (\fref{zzldos}(b)).
As seen from \fref{120ldos}(b), (c), and (g), 
there is no zero-energy peak at the sites near the corner
and hence edge states locally disappear.
However, edge states appear at the sites away from the corner.
Indeed we observe a broad peak at the site 17 (\fref{120ldos}(d)),
and the LDOS at the site 18 (\fref{120ldos}(e)) shows a sharp peak.

\begin{figure}[t]
	\begin{center}
		(a)\\
		\includegraphics[scale=0.6]{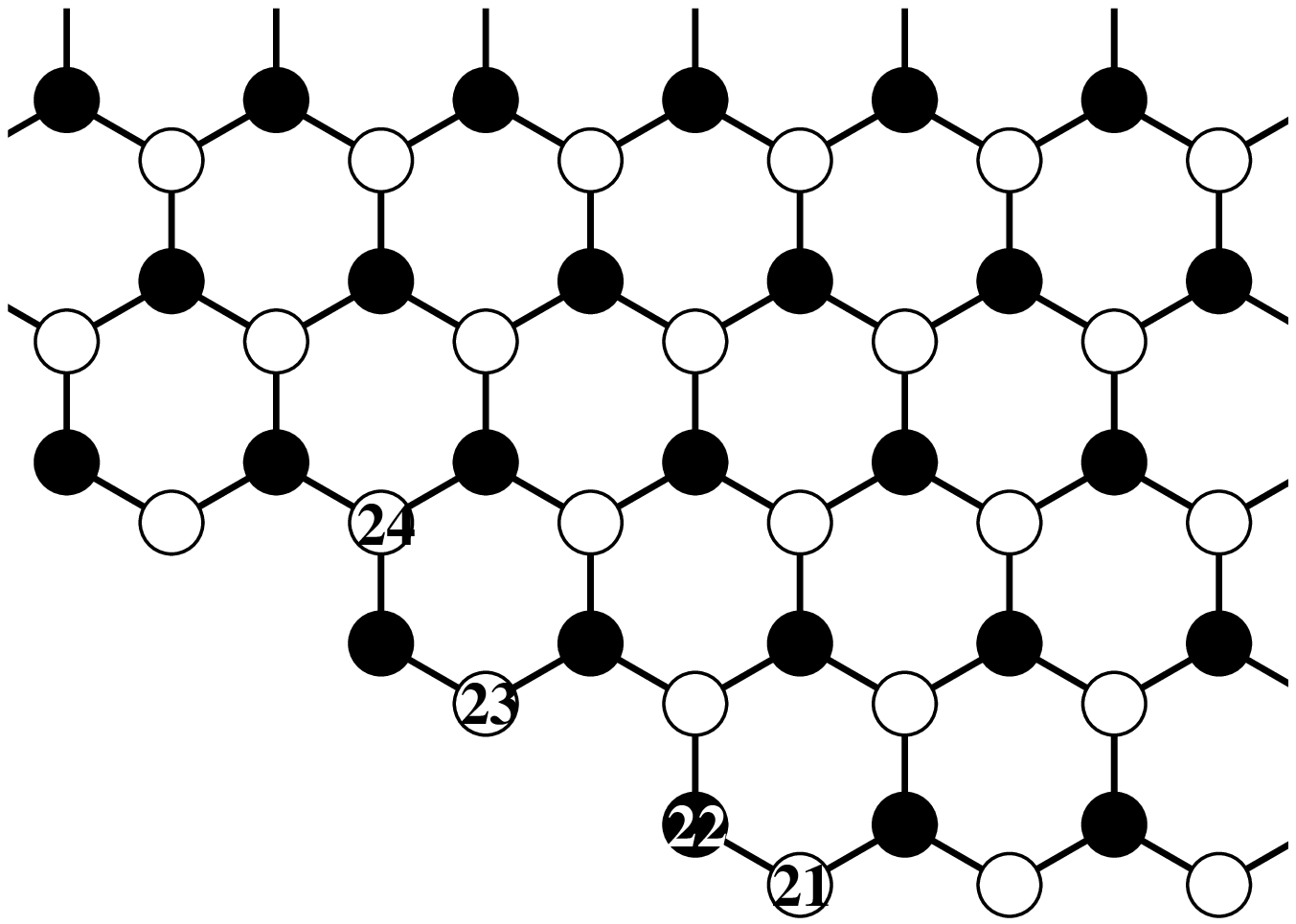}
	\end{center}
	\begin{minipage}[t]{0.495\textwidth}
		\begin{center}
			(b)\\
			21\\
			\includegraphics[scale=0.17]
			{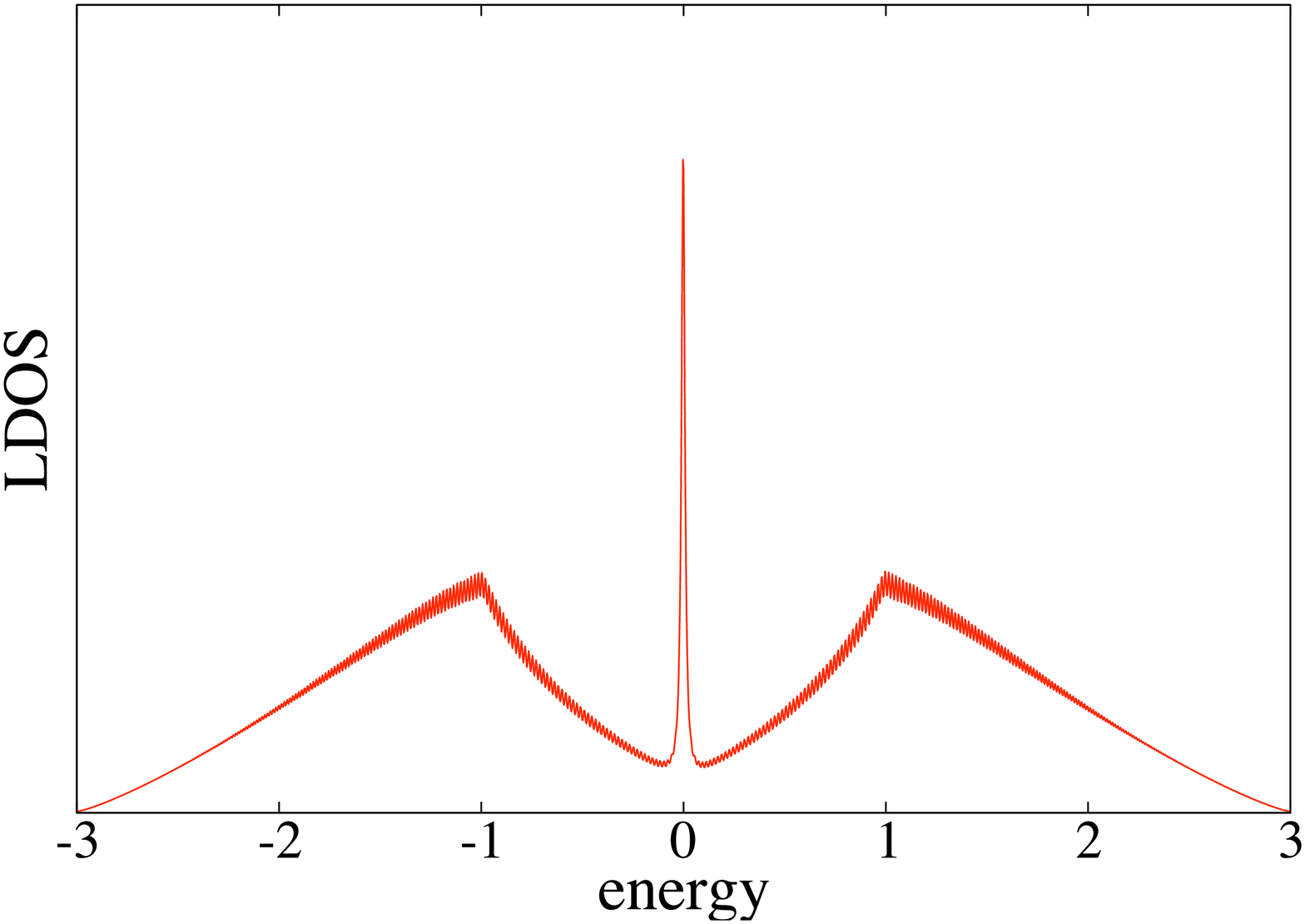}\\
		\end{center}
	\end{minipage}
	\begin{minipage}[t]{0.495\textwidth}
		\begin{center}
			(c)\\
			22\\
			\includegraphics[scale=0.17]
			{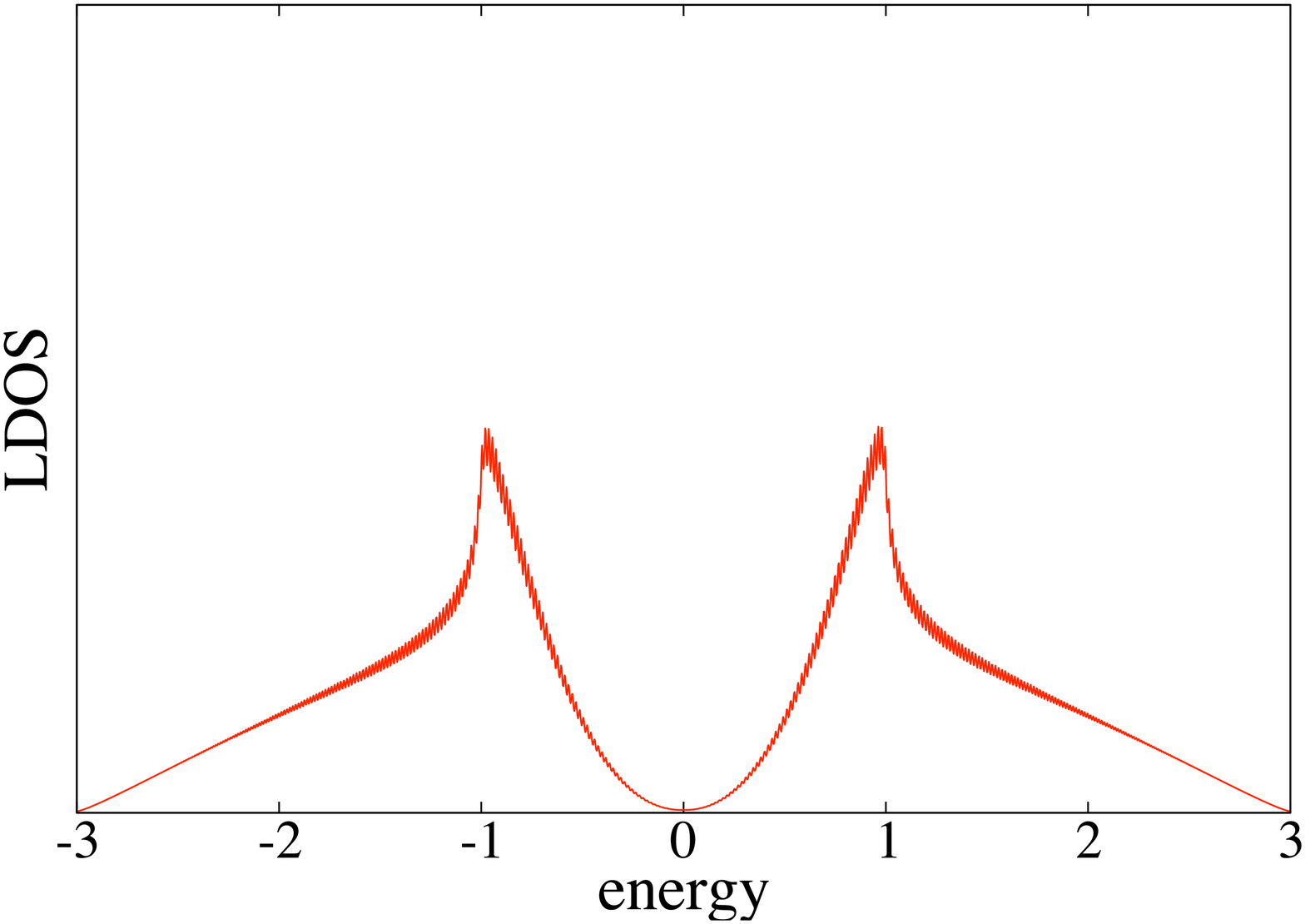}\\
		\end{center}
	\end{minipage}

\vspace{0.5cm}
	\begin{minipage}[t]{0.495\textwidth}
		\begin{center}
			(d)\\
			23\\
			\includegraphics[scale=0.17]
			{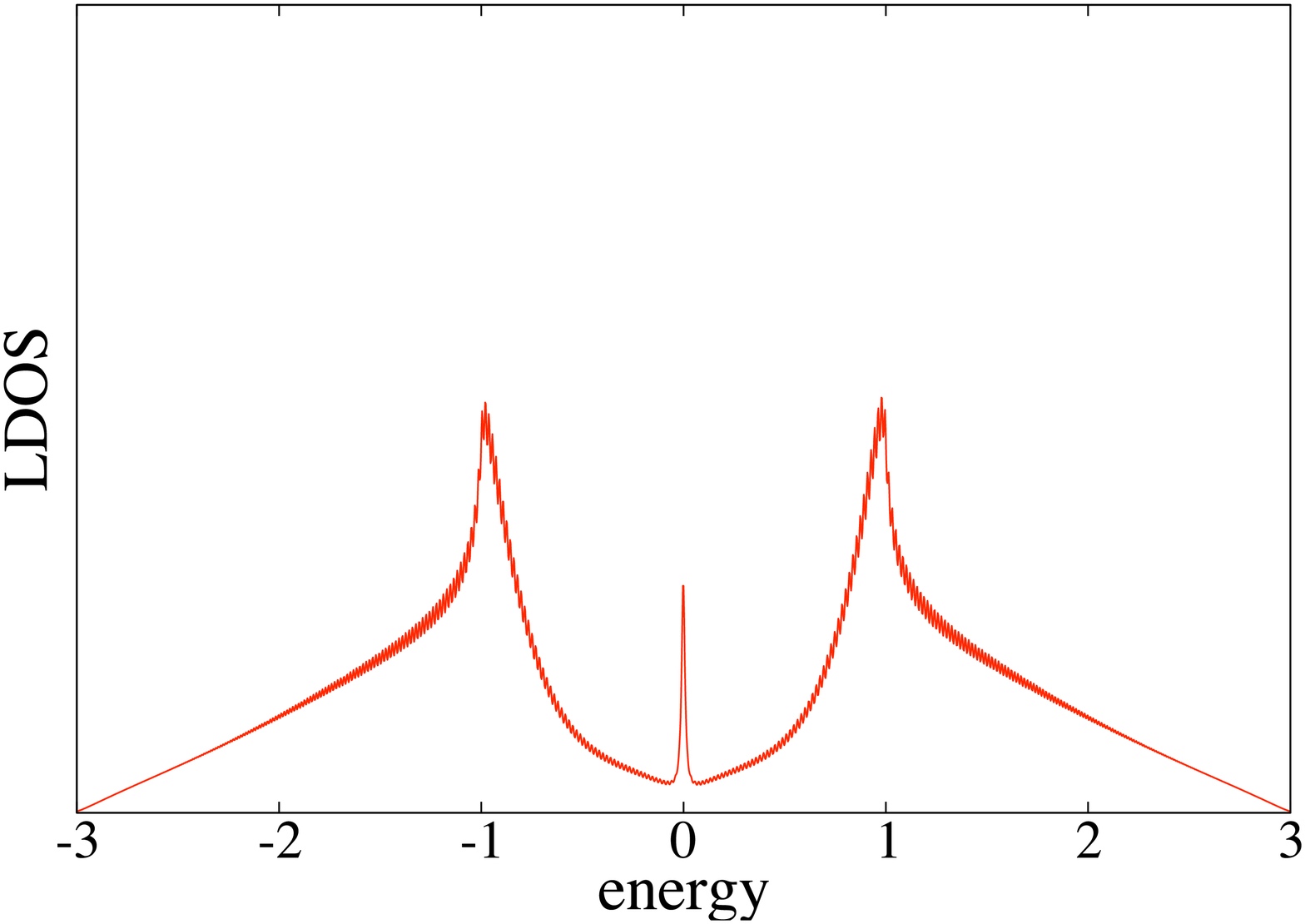}\\
		\end{center}
	\end{minipage}
	\begin{minipage}[t]{0.495\textwidth}
		\begin{center}
			(e)\\
			24\\
			\includegraphics[scale=0.17]
			{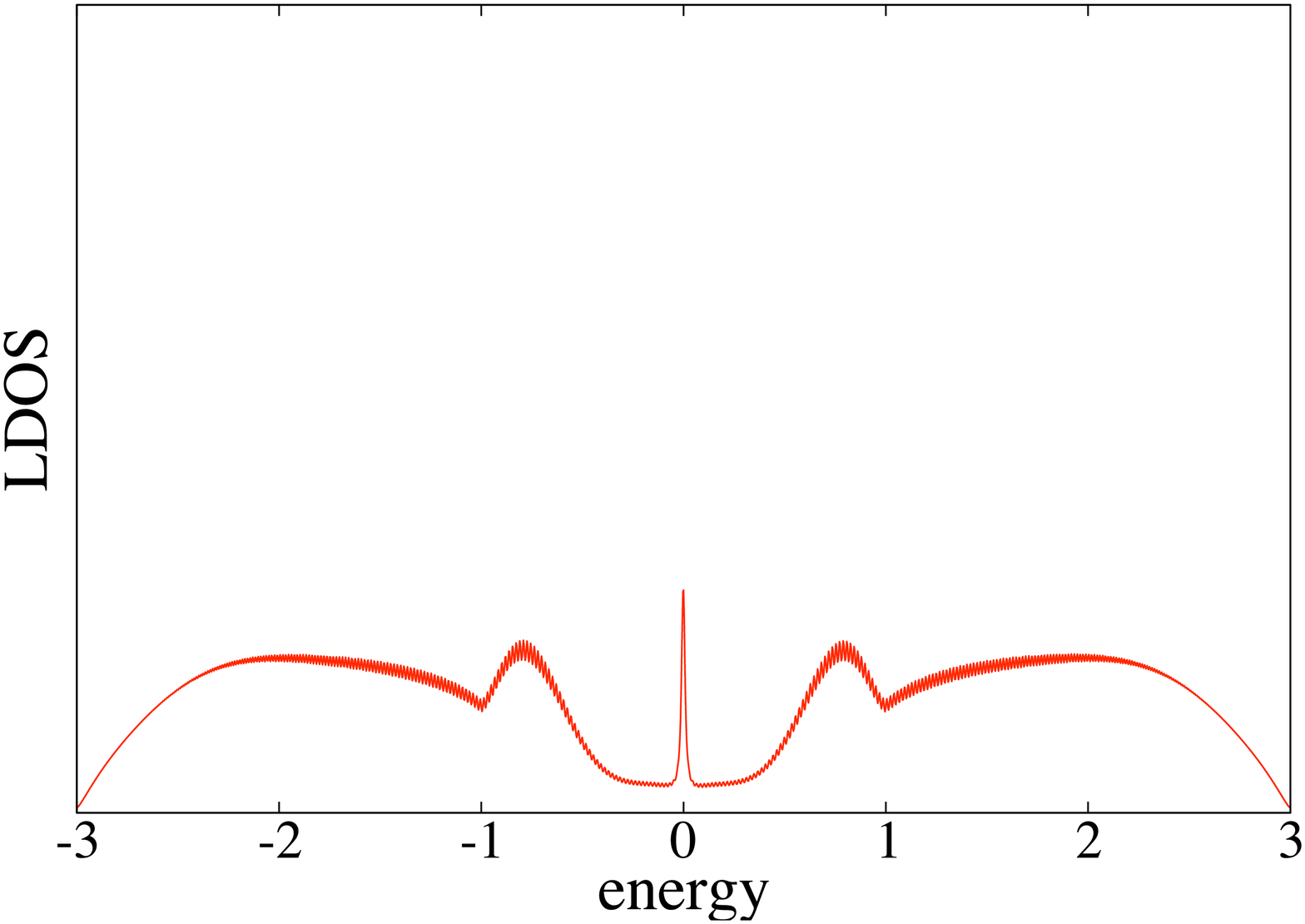}\\
		\end{center}
	\end{minipage}
	\caption{(a) The strcture of the 150$^\circ$ corner edge.
	(b), (c), (d), and (e) display the LDOS at the site 21, site 22, site 23, and site 24, respectively.
	The number indicated above each graph represents the site number defined in (a).
	}
	\label{150ldos}
\end{figure}

\noindent
\textit{{\rm (iv)} 150$^\circ$ corner edge.-}
Figure \ref{150ldos} shows the LDOS at several sites in the presence of the 150$^\circ$ corner edge.
As shown in \fref{150ldos}(b), (d), and (e), 
a zero-energy peak exists at the sites 21, 23 and 24 belonging to a same sublattice.
This indicates the existence of edge states.
As in the 90$^\circ$ case,
the LDOS near the corner possesses both the character of the LDOS in the single zz edge case and that in the single ac edge case.
For example, we can regard that the LDOS at the site 24 (\fref{150ldos}(e)) is a mixture of the LDOS at
the site 4 on the single zz edge (\fref{zzldos}(e)) and that at the site 6 on the single ac edge (\fref{acldos}(c)).
The peculiarity of the 150$^\circ$ case is that
the zero-energy peak of the site 21 is quite smaller than that at the site 1 on the single zz edge (\fref{zzldos}(b)).

\clearpage

\begin{figure}[tbp]
	\begin{minipage}{0.48\textwidth}
	\begin{center}
		\includegraphics[scale=0.5]{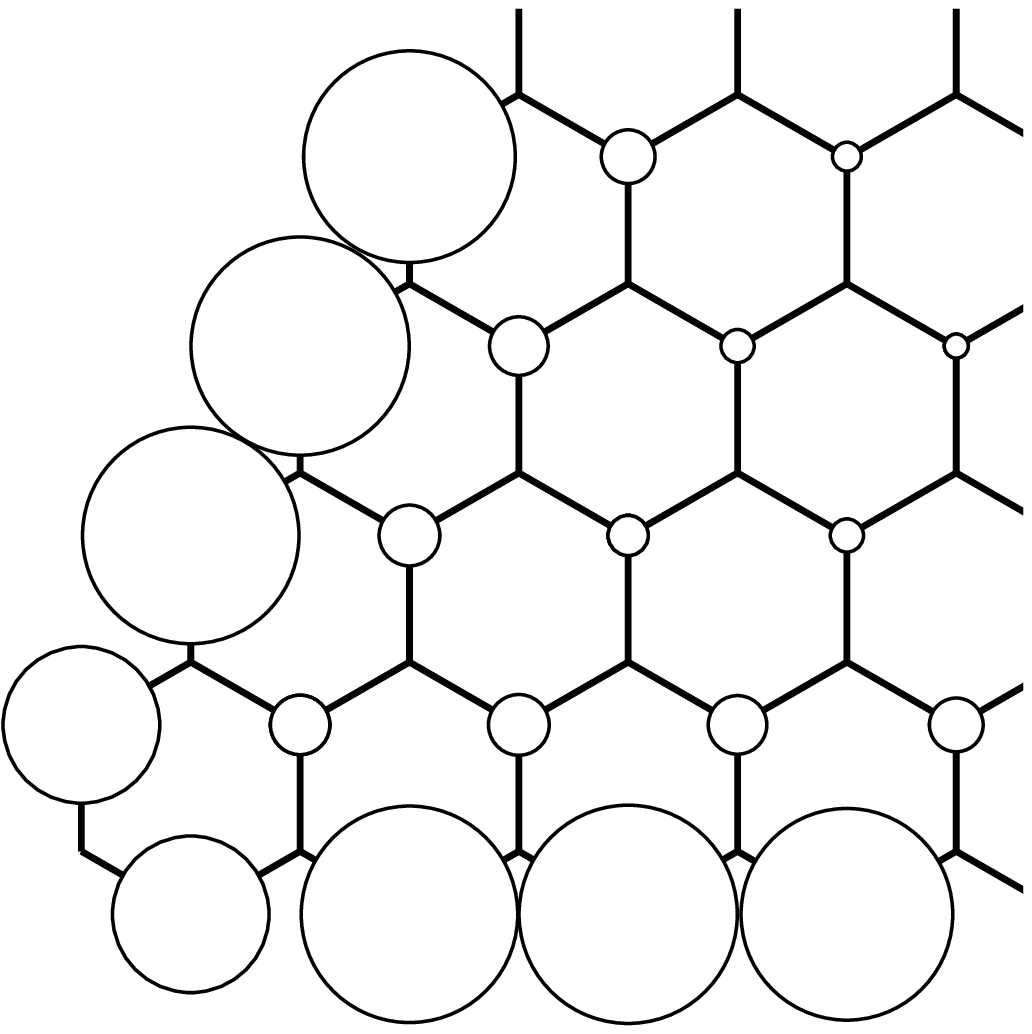}\\
		(a)
	\end{center}
	\end{minipage}
	\begin{minipage}{0.48\textwidth}
	\begin{center}
		\includegraphics[scale=0.5]{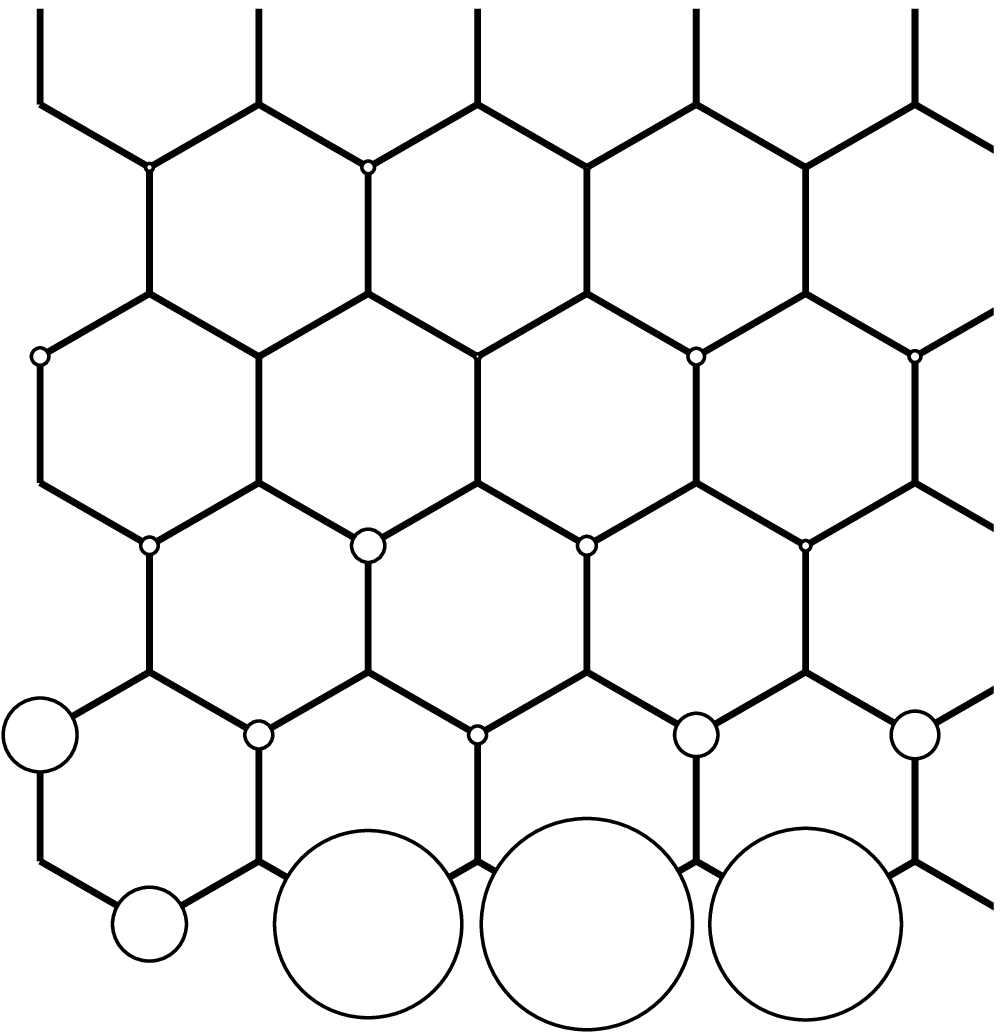}\\
		(b)
	\end{center}
	\end{minipage}\\
	\begin{center}
		\includegraphics[scale=0.5]{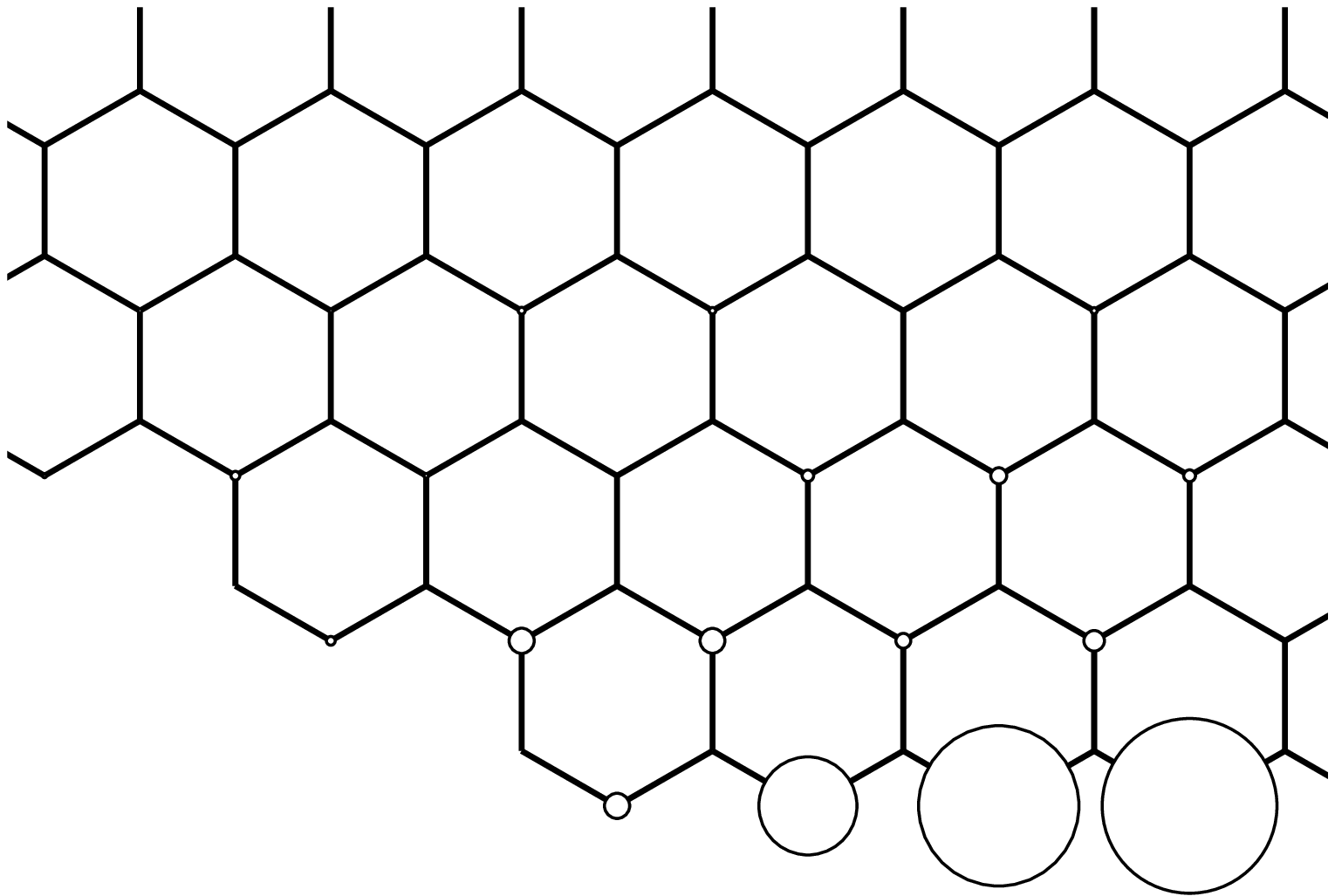}\\
		(c)
	\end{center}
	\caption{The LDOS in the presence of the (a) 60$^\circ$, (b) 90$^\circ$ 
	and (c) 150$^\circ$ corner edges at $\varepsilon=0$. 
	The radius of open circles indicates the magnitude of the LDOS.}
	\label{LDOS6090150}
\end{figure}

Figure \ref{LDOS6090150} represents the spatial dependance of the LDOS at $\varepsilon=0$
in the presence of the (a) 60$^\circ$, (b) 90$^\circ$, and (c) 150$^\circ$ corner edges.
A radius of each open circle indicates
the magnitude of the LDOS.
In these figures,
the LDOS has a finite value only on the sublattice involving zz edge sites
but vanishes on the other sublattice.
We observe that the LDOS localizes near zz edges,
indicating the presence of edge localized states.
However, special emphasis is placed on the case of the 60$^\circ$ corner edge,
where the magnitude of the LDOS at inner sites
is larger than that in the other two cases.
This reflects the fact that edge localized states are present 
at both the two zz edges.
The overlap of these edge localized states enhances the magnitude of the LDOS, i.e. constructive interference. 
\begin{figure}[t]
	\begin{center}
		\includegraphics[scale=0.5]{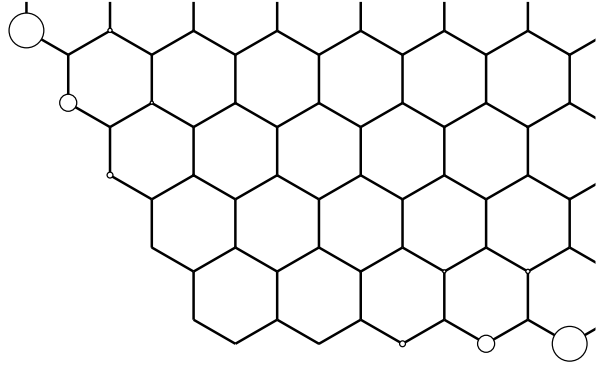}
	\end{center}
	\caption{The LDOS in the presence of the 120$^\circ$ corner edge at $\varepsilon=0$. 
	The radius of open circles indicates the magnitude of the LDOS.}
	\label{LDOS120}
\end{figure}
The other corner edge structures with the angle 90$^\circ$ or 150$^\circ$
consist of one zz edge and one ac edge.
Note the LDOS in the single ac edge system
vanishes at $\varepsilon=0$ on any sites.
In these corner edge structures,
the LDOS becomes finite even at $\varepsilon=0$ due to the presence of a zz edge.
Even at sites on the ac edge,
the LDOS can have a finite value.

Figure \ref{LDOS120} represents the spatial dependance of the LDOS  at $\varepsilon=0$
in the presence of the 120$^\circ$ corner edge.
In this figure, we observe that the LDOS vanishes at the sites near the corner,
indicating local disappearance of edge states, i.e. destructive interference.
In spite of the fact that the 120$^\circ$ corner edge consists of two zz edges,
the edge states are not fully stabilized in contrast to the case of the 60$^\circ$ corner edge.
Note that in the 120$^\circ$ case, 
edge sites on one zz edge and those on the other zz edge 
belong to different sublattices,
while all edge sites in the 60$^\circ$ case belong to a same sublattice.
As we discuss in the next section, this is the reason for the qualitative difference between the two cases.

\section{Analytical Treatment}
\begin{figure}[t]
	\begin{center}
		\includegraphics[scale=0.7]{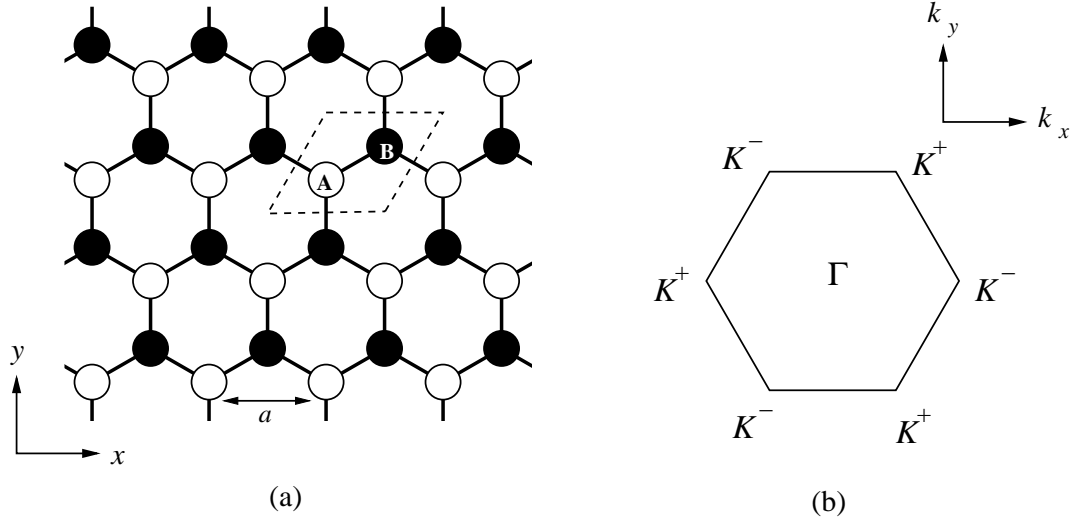}
	\end{center}
	\caption{(a) Honeycomb structure of graphene.
	The region enclosed in a broken line is a unit cell.
	A and B are non-equivalent sites which form sublattices.
	(b) The first Brillouin zone of graphene,
	where $\kp$, $\km$ and $\Gamma$ are symmetric points.
	The $K^+$ point is located 
	at $(-4\pi/3a,0)$ and $(2\pi/3a,\pm 2\pi/\sqrt{3}a)$
	while the $K^-$ point is located 
	at $(4\pi/3a,0)$ and $(-2\pi/3a,\pm 2\pi/\sqrt{3}a)$.
	}
	\label{ucbz}
\end{figure}
Our study on the LDOS reveals that edge localized states are stabilized 
in corner edge structures except for the 120$^\circ$ case.
To provide insight into this behavior, we analyze edge localized states in corner edge structures
by using an effective mass description,
which is applicable to low-energy states in the vicinity of the $\kpm$ point.
The $\kp$ and $\km$ points are characterized by 
$\bkp=(-4\pi/3a,0)$ and
$\bkm=(4\pi/3a,0)$, respectively.
Here, $a$ is lattice constant.
As shown in \fref{ucbz} (a), the unit cell of graphene
has two non-equivalent carbon atoms
A and B which form
A sublattice and B sublattice, respectively.
We represent
the wave function $\psia$ for A sublattice and 
the wave function $\psib$ for B sublattice
as
\begin{eqnarray}
	\psi_{\rm A}(\rr)={\rm e}^{{\rm i}\bkp\cdot\rr}\Fap(\rr)+{\rm e}^{{\rm i}\bkm\cdot\rr}\Fam(\rr),\\
	\psi_{\rm B}(\rr)={\rm e}^{{\rm i}\bkp\cdot\rr}\Fbm(\rr)-{\rm e}^{{\rm i}\bkm\cdot\rr}\Fbm(\rr),
	\label{efgattai}
\end{eqnarray}
where $F^{\pm}$ are envelope functions near the $\kpm$ point.
The envelope functions at energy $\varepsilon$ satisfy
\begin{equation}
	\gamma
	\begin{pmatrix}
		0&\hat{k}_x-{\rm i}\hat{k}_y&0&0\\
		\hat{k}_x+{\rm i}\hat{k}_y&0&0&0\\
		0&0&0&\hat{k}_x+{\rm i}\hat{k}_y\\
		0&0&\hat{k}_x-{\rm i}\hat{k}_y&0
	\end{pmatrix}
	\begin{pmatrix}
		F^+_{\rm A}({\boldsymbol r})\\F^+_{\rm B}({\boldsymbol r})\\F^-_{\rm A}({\boldsymbol r})\\F^-_{\rm B}({\boldsymbol r})
	\end{pmatrix}
	=
	\varepsilon	
	\begin{pmatrix}
		F^+_{\rm A}({\boldsymbol r})\\F^+_{\rm B}({\boldsymbol r})\\F^-_{\rm A}({\boldsymbol r})\\F^-_{\rm B}({\boldsymbol r})
	\end{pmatrix},
	\label{kdp}
\end{equation}
where $\gamma$ is a band parameter,
$\hat{k}_x=-{\rm i}\partial / \partial x$, and
$\hat{k}_y=-{\rm i}\partial /\partial y$.
This is called $\kdotp$ equation\cite{mcclure,Weiss}, which is an effective mass equation
for graphene systems.

For later convenience,
we present the envelope functions for edge states at $\varepsilon=0$\cite{wakad}.
Let us consider a semi-infinite graphene which occupies the region of $y>0$,
and has a zigzag edge at $y=0$.
Assumng that edge sites belong to A sublattice,
we adopt the boundary condition of $\Fbpm({\boldsymbol r})|_{y=0}=0$.
The envelope functions for edge states 
are given as 
\begin{equation}
	\begin{pmatrix}
		\Fapm({\boldsymbol r})\\ \Fbpm({\boldsymbol r})
	\end{pmatrix}
	=C
	\begin{pmatrix}
		{\rm e}^{\pm {\rm i}k_xx}{\rm e}^{-k_xy}\\0
	\end{pmatrix}
	,
	\label{kdpzz}
\end{equation}
where $C$ is a normalization constant.
The absolute value of $\Fapm$ in eq. \eref{kdpzz} has a maximum value at $y=0$ and 
exponentially decays with increaseing $y$.
This represents edge states localized along the zz edge.
We construct zero-energy wave functions 
which satisfy the boundary condition of corner edges by using
eq. \eref{kdpzz}.

\begin{figure}[t]
	\begin{center}
		\includegraphics[scale=0.5]{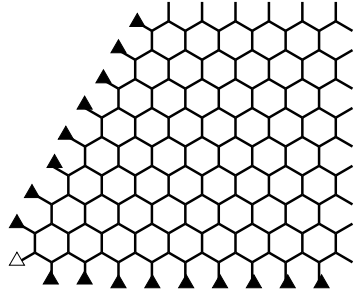}
	\end{center}
	\caption{The boundary condition	for the 60$^\circ$ corner edge
	requires that wave functions vanish
	at sites marked with a triangle.
	A site marked with an open triangle belongs to A sublattice, 
	and
	sites marked with a filled triangle belongs to B sublattice.
	}
	\label{60bound}
\end{figure}

\noindent
\textit{{\rm (i)} 60$^\circ$ corner edge.-}
We first consider wave functions in the presence of the 60$^\circ$ corner edge as shown in \fref{60bound}.
We attempt to construct wave functions near the $\kp$ point
in terms of two edge localized wave functions.
One is the wave function for the 0$^\circ$ zz edge
\begin{align}
	C
	\begin{pmatrix}
		{\rm e}^{-{\rm i}Kx}{\rm e}^{ {\rm i}k_x(x+{\rm i}y)}\\0
	\end{pmatrix},
	\label{houraku1}
\end{align}
and the other is the wave function for the 60$^\circ$ zz edge
\begin{align}
	C
	\begin{pmatrix}
		{\rm e}^{-{\rm i}Kx}{\rm e}^{ {\rm i}k_x(x+{\rm i}y){\rm e}^{ {\rm i}\frac{2}{3}\pi}}\\0
	\end{pmatrix},
	\label{houraku2}
\end{align}
where $K\equiv4\pi/3a$.
Here and hereafter we refer to zz edge intersecting the $x$ axis with angle $\theta$ degree as $\theta^\circ$ zz edge.
We adopt
their linear combination
\begin{equation}
	\begin{pmatrix}
		\psia\\\psib
	\end{pmatrix}=
	\begin{pmatrix} 
		{\rm e}^{-{\rm i}Kx}(C_1{\rm e}^{ {\rm i}k_x(x+{\rm i}y)}
		+C_2{\rm e}^{ {\rm i}k_x(x+{\rm i}y){\rm e}^{ {\rm i}\frac{2}{3}\pi}})\\
		0
        \end{pmatrix}
	\label{60wfproto}
\end{equation}
as a trial wave function
in the presence of the 60$^\circ$ corner edge.
The boundary condition 
requires that the wave function vanishes at sites 
marked with triangles in \fref{60bound}.
Because $\psib=0$,
we need to consider only the boundary condition for $\psia$.
Only the site at the corner with an open triangle
belongs to A sulattice.
We define this site as the origin of the coordinate.
Hence, the boundary condition for $\psia$ is simply given by
\begin{equation}
	\psi_{\rm A}(0,0)=0,
	\label{60bc}
\end{equation}
yielding $C_2=-C_1$.
We obtain the wave function in the presence of the 60$^\circ$ corner edge as
\begin{equation}
	\begin{pmatrix}
		\psia\\\psib
	\end{pmatrix}=
	C
	\begin{pmatrix} 
		{\rm e}^{-{\rm i}Kx}({\rm e}^{ {\rm i}k_x(x+{\rm i}y)}
		-{\rm e}^{ {\rm i}k_x(x+{\rm i}y){\rm e}^{ {\rm i}\frac{2}{3}\pi}})\\
		0
        \end{pmatrix}.
	\label{60wf}
\end{equation}
This indicates the existence of edge states in the 60$^\circ$ corner edge\cite{ezawa2}.

\begin{figure}[t]
	\begin{center}
		\includegraphics[scale=0.5]{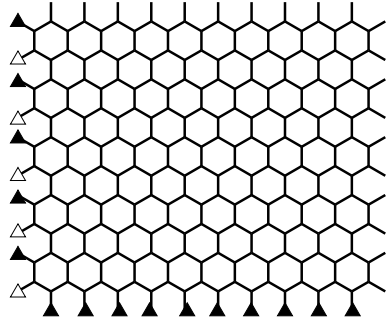}
	\end{center}
	\caption{The boundary condition	for 90$^\circ$ corner edge
	requires that wave functions vanish
	at sites marked with a triangle.
	Sites marked with an open triangle belongs to A sublattice, 
	and
	sites marked with a filled triangle belongs to B sublattice.
	}
	\label{90bound}
\end{figure}

\noindent
\textit{{\rm (ii)} 90$^\circ$ corner edge.-}
Secondly we consider the 90$^\circ$ corner edge as shown in \fref{90bound}.
In this case,
states near $\kp$ and $\km$ points are mixed due to the presence of an ac edge.
We construct zero-energy wave functions by using edge localized wave function
near the $\kp$ point,
\begin{align}
	&
	C
	\begin{pmatrix}
		{\rm e}^{-{\rm i}Kx}{\rm e}^{ {\rm i}k_x(x+{\rm i}y)}\\0
	\end{pmatrix},
	\label{houraku90}
\end{align}
and that near the $\km$ point,
\begin{align}
	&
	C
	\begin{pmatrix}
		{\rm e}^{ {\rm i}Kx}{\rm e}^{-{\rm i}k_x(x-{\rm i}y)}\\0
	\end{pmatrix}.
	\label{houraku91}
\end{align}
We adopt their linear combination
\begin{align}
	\begin{pmatrix}
	\psi_{\rm A}(\rr)\\\psi_{\rm B}(\rr)
	\end{pmatrix}
	=
	\begin{pmatrix}
		C_3{\rm e}^{-{\rm i}Kx}{\rm e}^{ {\rm i}k_x(x+{\rm i}y)}
		+C_4{\rm e}^{ {\rm i}Kx}{\rm e}^{-{\rm i}k_x(x-{\rm i}y)}\\0
	\end{pmatrix}
	\label{90sol}
\end{align}
as a trial wave function.
This must vanishes at sites marked with triangles in
\fref{90bound}.
Because $\psib=0$,
we need to consider only the boundary condition for $\psia$.
The sites marked with open triangles
belong to A sulattice.
We define the site at the corner with an open triangle as the origin.
The coordinates of the open triangles are $(x,y)=(0,\sqrt{3}a\times m)$
with $m=0,1,2,\ldots$.
Hence, the boundary condition for $\psia$ reads
\begin{equation}
	\psi_{\rm A}(0,\sqrt{3}a\times m)=0
	\;\;\;\;\;(m=0,1,2,\ldots).
	\label{90bc}
\end{equation}
Imposing this condition to $\psi_{\rm A}$ in eq. \eref{90sol},
we obtain $C_4=-C_3$.
We obtain the wave function for the 90$^\circ$ corner edge as
\begin{equation}
	\begin{pmatrix}
                \psia\\\psib
        \end{pmatrix}=
	C
	\begin{pmatrix} 
		{\rm e}^{-{\rm i}Kx}{\rm e}^{ {\rm i}k_x(x+{\rm i}y)} 
		-{\rm e}^{ {\rm i}Kx}{\rm e}^{-{\rm i}k_x(x-{\rm i}y)}\\
		0
        \end{pmatrix}.
	\label{90wf}
\end{equation}
This indicates the existence of edge states in the 90$^\circ$ corner edge.

\begin{figure}[t]
	\begin{center}
		\includegraphics[scale=0.5]{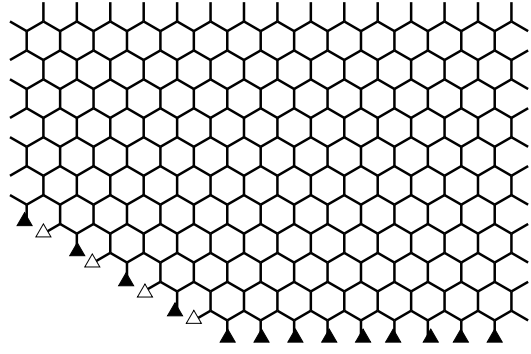}
	\end{center}
	\caption{The boundary condition	for the 150$^\circ$ corner edge
	requires that wave functions vanish
	at sites marked with a triangle.
	Sites marked with an open triangle belongs to A sublattice, 
	and
	sites marked with a filled triangle belongs to B sublattice.
	}
	\label{150bound}
\end{figure}

\noindent
\textit{{\rm (iii)} 150$^\circ$ corner edge.-}
Thirdly we consider the 150$^\circ$ corner edge as shown in \fref{150bound}.
We obtain zero-energy wave functions using a conformal mapping technique\cite{mapping}.
In terms of the complex variable $z\equiv x+iy$, eq. \eref{kdpzz} is rewritten as
\begin{equation}
	\begin{pmatrix}
		F^+_{\rm A}(z)\\F^+_{\rm B}(z)\\F^-_{\rm A}(z)\\F^-_{\rm B}(z)
	\end{pmatrix}
	=
	\begin{pmatrix}
		C{\rm e}^{ {\rm i}k_xz}\\0\\C'{\rm e}^{-{\rm i}k_xz^*}\\0
	\end{pmatrix},
	\label{wsol}
\end{equation}
where $z^*$ is the complex conjugate of $z$.
Here we introduce the transformation of $w=z^{3/5}$.
This transformation maps a 150$^\circ$ corner on $z$ plane 
to a 90$^\circ$ corner on $w$ plane and vice versa.
Thus, the wave function for the 90$^\circ$ corner edge on $w$ plane
\begin{align}
	\begin{pmatrix}
	\psi_{\rm A}(w)\\ 
	\psi_{\rm B}(w)
	\end{pmatrix}
	&=C
	\begin{pmatrix}
		{\rm e}^{-{\rm i}Kx}{\rm e}^{ {\rm i}k_xw}
		-{\rm e}^{ {\rm i}Kx}{\rm e}^{-{\rm i}kw^*}\\0
	\end{pmatrix}
	\label{150wsolf}
\end{align}
is mapped to
\begin{align}
	\begin{pmatrix}
	\psi_{\rm A}(\rr)\\
	\psi_{\rm B}(\rr)
	\end{pmatrix}
	&=C
	\begin{pmatrix}
		{\rm e}^{-{\rm i}Kx}{\rm e}^{ {\rm i}k_xz^{\frac{3}{5}}}
		-{\rm e}^{ {\rm i}Kx}{\rm e}^{-{\rm i}k_xz^{*\frac{3}{5}}}\\
	0
	\end{pmatrix}\\
	&=C
	\begin{pmatrix}
		{\rm e}^{-{\rm i}Kx}{\rm e}^{ {\rm i}k_x(x+{\rm i}y)^{\frac{3}{5}}}
		-{\rm e}^{ {\rm i}Kx}{\rm e}^{-{\rm i}k_x(x-{\rm i}y)^{\frac{3}{5}}}\\
	0
	\end{pmatrix}
	\label{150zsolf}
\end{align}
on $z$ plane.
The boundary condition requires that the wave function vanishes
at sites marked with triangles in \fref{150bound}.
Again, we need to consider only the boundary condition for $\psia$.
The sites marked with open triangles
belong to A sulattice.
We define the site at the corner with an open triangle as the origin.
The coordinates of the open triangles are $(x,y)=(-\frac{3}{2}a\times m,\frac{\sqrt{3}}{2}a\times m)$
with $m=0,1,2,\ldots$.
Hence, the boundary condition for $\psia$ is given by
\begin{equation}
	\psi_{\rm A}(-\frac{3}{2}a\times m,\frac{\sqrt{3}}{2}a\times m)=0
	\;\;\;\;\;(m=0,1,2,\ldots).
	\label{150bc}
\end{equation}
The wave funciton $\psi_{\rm A}$ in eq. \eref{150zsolf} satisfies this condition.
Therefore eq. \eref{150zsolf} can be considered as a wave function
in the presence of the 150$^\circ$ corner edge.
This indicates the existence of edge states.
The envelope funcions ${\rm e}^{ {\rm i}k_x(x+{\rm i}y)^{\frac{3}{5}}}$ and
${\rm e}^{-{\rm i}k_x(x-{\rm i}y)^{\frac{3}{5}}}$ 
are diffrent from ordinary envelope functions given in eq. \eref{kdpzz},
but both satisfy the $\kdotp$ equation given in eq. \eref{kdp}.

\begin{figure}[t]
	\begin{center}
		\includegraphics[scale=0.5]{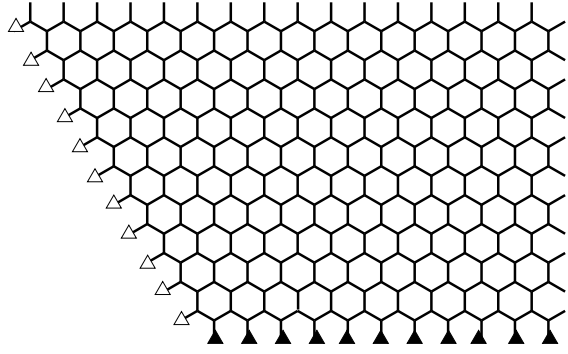}
	\end{center}
	\caption{The boundary condition	for the 120$^\circ$ corner edge
	requires that wave functions vanish
	at sites marked with a triangle.
	Sites marked with an open triangle belongs to A sublattice, 
	and
	sites marked with a filled triangle belongs to B sublattice.
	}
	\label{120bound}
\end{figure}

\noindent
\textit{{\rm (iv)} 120$^\circ$ corner edge.-}
Lastly we consider the 120$^\circ$ corner edge as shown in \fref{120bound}.
The boundary condition requires that
$\psib$ vanishes
at sites marked with a filled triangle
and $\psia$ vanishes 
at sites marked with an open triangle.
Therefore, both $\psia$ and $\psib$
are subjected to the boundary condition,
in contrast to the 60$^\circ$ case
where one component is free from the boundary condition.
This crucially affects zero-energy states in the 120$^\circ$ case as we see below.
Similar to the treatment for the 60$^\circ$ case,
we first adopt a linear combination of the wave function for
the 0$^\circ$ zz edge and that for the 120$^\circ$ zz edge
as a trial wave function.
The wave function for the 0$^\circ$ zz edge near the $\kp$ point is
\begin{align}
	&
	C
	\begin{pmatrix}
		{\rm e}^{-{\rm i}Kx}{\rm e}^{ {\rm i}k_x(x+{\rm i}y)}\\0
	\end{pmatrix}
	\label{houraku902}
\end{align}
and the wave function for the 120$^\circ$ zz edge near the $\kp$ point is
\begin{align}
	C
	\begin{pmatrix}
		0\\{\rm e}^{-{\rm i}Kx}{\rm e}^{-{\rm i}k_x(x-{\rm i}y){\rm e}^{-{\rm i}\frac{\pi}{3}}}
	\end{pmatrix}.
	\label{120zz}
\end{align}
The former
has only the A-sublattice component,
while the latter has only the B-sublattice component.
Obviously, their linear combination does
not satisfy the boundary condition
for both the A-sublattice and B-sulattice components.
We next consider a linear combination of the 0$^\circ$ edge wave functions
near the $\kp$ and $\km$ points,
\begin{align}
	\begin{pmatrix}
	\psi_{\rm A}(\rr)\\\psi_{\rm B}(\rr)
	\end{pmatrix}
	=
	\begin{pmatrix}
		C_5{\rm e}^{-{\rm i}Kx}{\rm e}^{ {\rm i}k_x(x+{\rm i}y)}
		+C_6{\rm e}^{ {\rm i}Kx}{\rm e}^{-{\rm i}k_x(x-{\rm i}y)}\\0
	\end{pmatrix}.
	\label{90sol2}
\end{align}
This is equivalent to eq. \eref{90sol}.
Though $\psi_{\rm B}$ of eq. \eref{90sol2} satisfies
the boundary condition,
$\psi_{\rm A}$ cannot satisfy the boundary condition
for arbitrary $C_5$ and $C_6$.
Finally, we consider a linear combination of 
the 0$^\circ$ zz edge wave function and
an arbitrary evanescent wave function near the $\kpm$ point given by
\begin{align}
	C
	\begin{pmatrix}
		{\rm e}^{\mp {\rm i}Kx}{\rm e}^{\pm {\rm i}p(x\pm {\rm i}y){\rm e}^{\pm {\rm i}\theta}}
		\\0
	\end{pmatrix}.
	\label{damping}
\end{align}
This wave function, reducing to the 0$^\circ$ zz edge wave function when
$\theta\rightarrow0$, satisfies eq. \eref{kdp} and
is bounded for $0\leq\theta\leq\frac{\pi}{3}$ 
in the 120$^\circ$ case.
Their linear combination
\begin{align}
	\begin{pmatrix}
	\psi_{\rm A}(\rr)\\\psi_{\rm B}(\rr)
	\end{pmatrix}
	=
	\begin{pmatrix}
		C_7{\rm e}^{-{\rm i}Kx}{\rm e}^{ {\rm i}k_x(x+{\rm i}y)}
		+C_8{\rm e}^{\mp {\rm i}Kx}{\rm e}^{\pm {\rm i}p(x\pm {\rm i}y)e^{\pm {\rm i}\theta}}\\0
	\end{pmatrix}.
	\label{120notsol}
\end{align}
does not satisfy the A-sublattice boundary condition for arbitrary
$C_7$, $C_8$, $p$, and $\theta$ as long as $p$ is sufficiently small.

We failed to construct zero-energy wave functions 
in the 120$^\circ$ case
in the form of a linear combination of the edge states, 
in striking contrast to the 60$^\circ$ case.
It is considered that this corresponds to 
the disappearance of the LDOS peak at $\varepsilon=0$ near the corner
observed in the numerical result.
We suppose that correct zero-energy states consist of
zz edge states
and complex scattered waves.
We point out that the sublattice configuration of two zz edges
plays a crucial role in the qualitative difference
between the 60$^\circ$ and 120$^\circ$ cases.

In the remaining of this section
we briefly consider the behavior of the LDOS shown in \fref{LDOS6090150},
on the basis of the wave functions obtained above.
Figure \ref{LDOS6090150} shows the spatial dependence of the LDOS at $\varepsilon=0$
in the 60$^\circ$, 90$^\circ$ and 150$^\circ$ cases.
We observe that the LDOS on a zz edge is slightly suppressed
in the close vicinity of a corner.
This should be distinguished from the strong suppression of 
the LDOS observed near a 120$^\circ$ corner,
and is simply accounted for on the basis of the wave functions
for zero-energy edge localized states presented in eqs. \eref{60wf}, \eref{90wf} and \eref{150zsolf}.
We see that due to destructive interference, the amplitude of
these wave functions is suppressed in the close vicinity of a corner
located at $(x,y)=(0,0)$ for a sufficiently small $k_x$.
This accounts for the slight suppression of the LDOS.
\section{Summary}

We have studied electronic states in semi-infinite graphene
with a corner edge, focusing on the stability of edge localized states.
The $60^{\circ}$, $90^{\circ}$, $120^{\circ}$ and $150^{\circ}$
corner edges are examined.
The $90^{\circ}$ and $150^{\circ}$ corner edges consist
of one zz edge and one ac edge, while the $60^{\circ}$ and $120^{\circ}$
corner edges consist of two zz edges.
We have numerically obtained the local density of states
on the basis of a nearest-neighbor tight-binding model
by using Haydock's recursion method.
We have shown that edge localized states appear along a zz edge
of each corner edge structure
except for the $120^{\circ}$ case.
In the $120^{\circ}$ case, we have also shown that edge localized states
locally disappear near the corner
but emerge with increasing the distance from the corner along each zz edge.
To provide insight into these behaviors,
we have analyzed electronic states at $\varepsilon = 0$
within the framework of an effective mass equation.
Except for the $120^{\circ}$ case, we have succeeded to obtain eigenstates
of the effective mass equation by forming a superposition of pair of
edge localized wave functions for an infinitely long straight zz edge.
This indicates the existence of edge localized states,
and is consistent with the behavior of the local density of states.
Contrastingly, no eigenstate has been obtained in such a simple form
in the $120^{\circ}$ case.
This suggests a possibility that the local disappearance of
edge localized states in the $120^{\circ}$ case
is beyond the effective mass description.
Note that although both the $60^{\circ}$ and $120^{\circ}$ corner edges
consist of two zz edges, zero-energy eigenstates of
the effective mass equation are obtained only in the former case.
We have pointed out that this reflects the fact
that two zz edges belong to a same sublattice in the former case
while they belong to different sublattices in the latter case.

\section*{Acknowledgment}

This work was supported in part by a Grant-in-Aid for Scientific
Research (C) (No. 21540389)
from the Japan Society for the Promotion of Science,
and by a Grant-in-Aid
for Specially promoted Research (No. 20001006)
from the Ministry of Education, Culture, Sports, Science and Technology.

\end{document}